\documentclass[11pt]{article}

\usepackage[margin=1in]{geometry}
\usepackage{amsmath,amssymb,amsthm}
\usepackage{booktabs}
\usepackage{graphicx}
\usepackage[authoryear,round,semicolon]{natbib}

\usepackage{microtype}
\usepackage{placeins}

\usepackage{algorithm}
\usepackage{algpseudocode}
\usepackage[hidelinks]{hyperref}

\graphicspath{{figures/}}

\makeatletter
\let\paper@makecaption\@makecaption
\long\def\@makecaption#1#2{%
  \begingroup
  \small
  \paper@makecaption{\textbf{#1}}{#2}%
  \endgroup
}
\makeatother

\makeatletter
\renewcommand{\@maketitle}{%
  \null
  \vskip 1em%
  \begin{center}%
    {\LARGE \@title \par}%
    \vskip 0.8em%
    {\large \@author \par}%
  \end{center}%
  \par
  \vskip 0.5em%
}
\makeatother

\newtheorem{theorem}{Theorem}

\newtheorem{corollary}{Corollary}

\newtheorem{assumption}{Assumption}

\newcommand{\source}{\mathrm{s}}
\newcommand{\target}{\mathrm{t}}

\title{High-Dimensional Nonsparse Regression under Covariate Shift via Importance-Weighted Partial Least Squares}
\author{\small
  Youwu Lin\textsuperscript{1},
  Jianghe Shi\textsuperscript{1},
  Pei Wang\textsuperscript{2}\thanks{Corresponding author, Email: pei.wang@gdufs.edu.cn}\\[6pt]
  \small \textsuperscript{1} School of Mathematics and Computing Science, \\
  \small Guilin University of Electronic Technology, 541002, Guilin, P.R.China\\
  \small \textsuperscript{2} School of Business, Guangdong University of Foreign Studies, 510006, Guangzhou, P.R.China
}
\date{}

\begin{document}

\maketitle

\begin{abstract}
This paper studies high-dimensional nonsparse regression when the data exhibit both strong predictor dependence and covariate shift. Many existing regression methods struggle with highly correlated predictors, while covariate shift can make components learned from source data less effective for target-domain prediction. To address these joint challenges, we develop an importance-weighted partial least squares (IWPLS) method that reweights source observations to reflect the target covariate distribution and extracts response-relevant components for coefficient estimation. The resulting estimator depends on weighted sample covariances, which generally lack the usual Wishart structure even when the target-domain covariates are Gaussian. We therefore combine a perturbation analysis of weighted PLS moments with an influence-function decomposition to establish consistency for the one- and multi-component estimators and asymptotic normality for the one-component estimator. Simulation studies and real-data analyses show that IWPLS outperforms the benchmark methods considered.
\end{abstract}

\section{Introduction}
With advances in data collection and storage, high-dimensional data with strongly correlated predictors have become common in chemometrics, economics, and genomics. In nonsparse regression, this dependence makes it difficult to estimate the coefficients of the many predictors that contribute to the response. Beyond this estimation difficulty, reliable prediction becomes more challenging across instruments, batches, time periods, or populations. In these settings, shifts in the covariate distribution between training and prediction can impair target-domain performance~\citep{ZechEtAl2018,NestorEtAl2019,LarsenEtAl2020,SinghMhasawadeChunara2022}. In this paper, we study high-dimensional nonsparse regression with strongly correlated predictors under covariate shift, assuming that the conditional distribution of the response given the predictors remains unchanged across the source and target domains.

We observe labeled source data and unlabeled target covariates, with a scalar response $Y$ and a predictor vector $\boldsymbol X\in\mathbb R^p$. Our theoretical analysis in Section~4 uses the linear model
\[
  Y^a=\alpha_0+(\boldsymbol X^a)^\top\boldsymbol\beta_0+\varepsilon^a,
  \qquad a\in\{\source,\target\}.
\]
Here, $\source$ and $\target$ index the source and target domains. The intercept $\alpha_0$ and the potentially dense coefficient vector $\boldsymbol\beta_0$ are shared across domains, while the predictor distributions may differ. The errors are independent of the covariates and share a common mean-zero distribution. Our goal is to use the two samples to construct a linear prediction rule for the target population.

In recent years, regression under covariate shift has attracted increasing attention. A major line of work is based on importance weighting, including weighted least squares~\citep{Shimodaira2000CovariateShift} and reweighted kernel ridge regression~\citep{MaPathakWainwright2023}. These methods derive their weights from the target-to-source covariate density ratio, which can be estimated directly~\citep{TsuboiEtAl2009}. In related high-dimensional transfer learning, \citet{ZhaoZhengLeng2026} introduced residual importance-weighted transfer learning for linear regression with a sparse target coefficient vector. These studies illustrate how weighting can adapt regression across domains, but they do not directly address response-guided dimension reduction when predictors are strongly correlated and the regression signal is nonsparse.

Partial least squares (PLS) provides one way to extract this structure by constructing components that reflect the relationship between predictors and responses~\citep{DeJong1993SIMPLS,WoldSjostromEriksson2001,CookHellandSu2013}. At the population level, the PLS construction is governed by the predictor covariance and the predictor--response cross-covariance, which together determine the response-relevant component structure~\citep{CookHellandSu2013}. Under covariate shift, however, components extracted from source data may be less effective for target-domain prediction, particularly when only a few components are retained. This limitation is one manifestation of a broader challenge in high-dimensional nonsparse regression with strongly correlated predictors and covariate shift.

\begin{figure}[htbp]
\centering
\includegraphics[width=\textwidth]{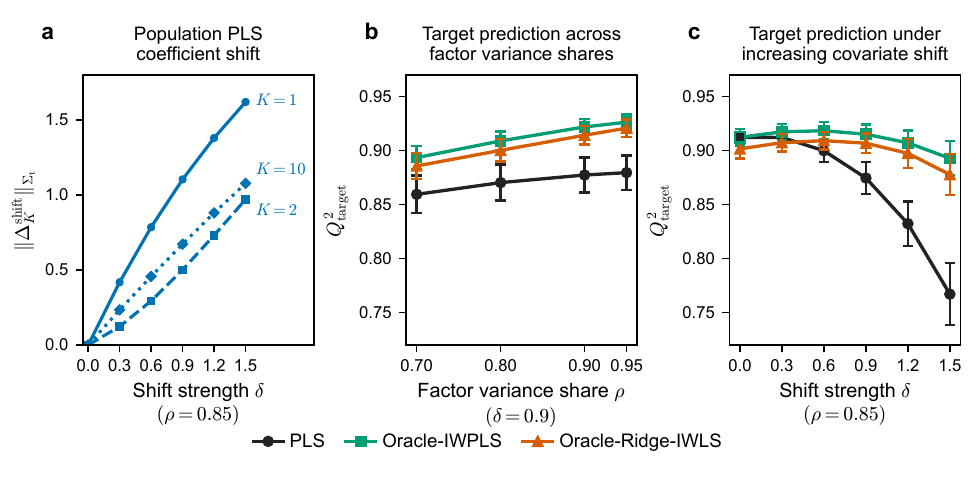}
\caption{
Predictor dependence and covariate shift in high-dimensional nonsparse regression. (a) Source--target difference in population PLS coefficients, measured in the target covariance norm, for $K\in\{1,2,10\}$ retained components. (b, c) Target-domain $Q^2$ as the source factor variance share $\rho$ and shift magnitude $\delta$ vary, respectively. The oracle methods use the true target-to-source density ratio. Points and error bars in (b) and (c) show means and standard deviations over 100 independent replications.
}
\label{fig:toy-example}
\end{figure}

To illustrate how predictor dependence and covariate shift affect target-domain prediction, we consider a toy example with $p=1000$ predictors and dense regression coefficients. We generate Gaussian source and target covariates whose means and covariance matrices depend on a shift parameter $\delta$, while keeping the conditional response model unchanged across domains. The response combines a dense linear signal with a quadratic term, as specified in Section~5. Each replication uses 800 labeled source observations, 800 unlabeled target observations, and an independent target-test sample of size 800. We vary the source factor variance share $\rho$ at $\delta=0.9$ and the shift magnitude $\delta$ at $\rho=0.85$, with 100 replications per setting. We compare PLS with oracle importance-weighted PLS (Oracle-IWPLS) and oracle importance-weighted ridge regression (Oracle-Ridge-IWLS). Both weighted methods use the true target-to-source density ratio. Figure~\ref{fig:toy-example} summarizes the source--target discrepancy in population PLS coefficients and target-domain $Q^2$. The source--target coefficient discrepancy increases with shift for all three component counts, $K\in\{1,2,10\}$ (panel a). Higher factor variance shares improve $Q^2$ for all three methods, but PLS remains below the two oracle methods (panel b). PLS and Oracle-IWPLS coincide without shift, and their prediction gap widens as the shift increases (panel c). These results show that source-domain PLS remains sensitive to covariate shift in this correlated, nonsparse setting.

These patterns motivate the development of importance-weighted partial least squares (IWPLS) for high-dimensional nonsparse regression under covariate shift. IWPLS uses target-to-source importance weights to construct predictor and predictor--response moments from labeled source observations. It then extracts response-relevant PLS components from these weighted moments and uses them to estimate the regression coefficients. In this way, both component extraction and coefficient estimation reflect the target covariate distribution. Importance weighting alters the stochastic behavior of the sample moments underlying PLS, so establishing the statistical properties of IWPLS requires new arguments beyond those developed for the unweighted estimator.

Even for unweighted PLS, rigorous high-dimensional analysis has proved challenging. \citet{helland1990partial} characterized PLS at the population level, and \citet{CookHellandSu2013} later connected its population structure to envelope methodology. \citet{CookForzani2018BigData} studied high-dimensional prediction for one-component PLS, while \citet{CookForzani2019} extended the analysis across signal regimes and identified favorable behavior in abundant regressions. More recently, \citet{BasaCookForzaniMarcos2024} derived an asymptotic distribution for a linear contrast of the one-component PLS coefficient estimator. These results inform our analysis, but they concern unweighted PLS within a single population.

The use of importance-weighted moments in IWPLS
introduces two theoretical challenges. With known
importance weights, the weighted sample covariance
generally lacks the usual Wishart structure, even
when target-domain predictors are Gaussian.
Estimating the density ratio introduces additional
perturbations to the moments underlying PLS.
To address these challenges, we control
self-normalized weighted moments and their
propagation through the normalized Krylov system.
We establish coefficient consistency for
fixed-component IWPLS, prediction-error bounds
for the corresponding prediction rules, and
projected asymptotic normality for the
one-component estimator. The distributional
analysis covers oracle weights and extends to
estimated weights under additional first-order
linearization and moment conditions. For correctly
specified fixed-basis uLSIF, we provide sufficient
conditions for these requirements that allow reuse
of the source covariates employed in IWPLS fitting.

The remainder of the paper is organized as follows. Section~2 describes the covariate-shift setting and defines the target-domain PLS functional. Section~3 presents IWPLS with oracle and estimated importance weights, together with practical considerations for density-ratio estimation and component selection. Section~4 establishes consistency for the one- and multi-component estimators and asymptotic normality for fixed-dimensional projections of the one-component coefficient estimator. Section~5 uses simulations to examine the effects of predictor dependence, covariate shift, and weight estimation. Section~6 reports analyses of the IDRC and Melamine transfer tasks. Section~7 concludes with a discussion of the method's scope and limitations. Proofs are provided in the Supplementary Material.

\FloatBarrier

\section{Problem Setup}
\subsection{Covariate-Shift Setting and Prediction Risk}

Let \(\boldsymbol{X}\in\mathbb R^p\) and \(Y\in\mathbb R\) denote the predictor vector and response. For \(a\in\{\source,\target\}\), let \(P_a\) denote the joint distribution of \((\boldsymbol{X},Y)\) in domain \(a\), with predictor marginal \(P_a^X\). We observe two independent samples: a labeled source sample
\[
  \mathcal D_{\source}
  =
  \left\{
    (\boldsymbol{X}_i^{\source},Y_i^{\source}) :
    i=1,\ldots,n_{\source}
  \right\},
  \qquad
  (\boldsymbol{X}_i^{\source},Y_i^{\source})
  \overset{\mathrm{i.i.d.}}{\sim}P_{\source},
\]
and an unlabeled target sample
\[
  \mathcal X_{\target}
  =
  \left\{
    \boldsymbol{X}_j^{\target} :
    j=1,\ldots,n_{\target}
  \right\},
  \qquad
  \boldsymbol{X}_j^{\target}
  \overset{\mathrm{i.i.d.}}{\sim}P_{\target}^{X}.
\]
Model construction and tuning use the labeled source observations and unlabeled target covariates. Target responses are reserved for predictive evaluation on separate target test samples, as described in Sections~5--6.

We assume covariate shift: the predictor marginals may differ across domains, while the conditional response distribution remains unchanged. We also require the target predictor distribution to be absolutely continuous with respect to the source predictor distribution:
\begin{equation}
\label{eq:covariate-shift-setting}
  P_{\source}(Y\in\cdot\mid \boldsymbol{X}=\boldsymbol{x})
  =
  P_{\target}(Y\in\cdot\mid \boldsymbol{X}=\boldsymbol{x})
  \quad\text{for }P_{\target}^{X}\text{-almost every }\boldsymbol{x},
  \qquad
  P_{\target}^{X}\ll P_{\source}^{X}.
\end{equation}
Absolute continuity ensures that the target-to-source density ratio is well defined. Together with conditional invariance, it allows target-domain expectations to be written as importance-weighted source-domain expectations.

Throughout, \(\|\boldsymbol v\|_2\) denotes the Euclidean norm, while \(\|\mathbf A\|_{\mathrm{op}}\) and \(\|\mathbf A\|_{\mathrm F}\) denote the operator and Frobenius norms. For an integrable function \(f\), write \(P_a^X f=\int f(\boldsymbol x)\,dP_a^X(\boldsymbol x)\). When \(f\) is fitted from data, this integral holds the fitted function fixed.

For \(a\in\{\source,\target\}\), let \(\mathbb{E}_a\) denote expectation under \(P_a\). We assume finite second moments,
\[
  \mathbb{E}_a\|\boldsymbol{X}\|_2^2<\infty,
  \qquad
  \mathbb{E}_a(Y^2)<\infty,
\]
so that the means, covariance matrices, and predictor--response cross-covariances used below are finite. The additional signal, weighted-moment, and stability conditions for high-dimensional analysis are specified in Section~4.

Write
\[
  \boldsymbol{\mu}_a=\mathbb{E}_a(\boldsymbol{X}),
  \qquad
  \nu_a=\mathbb{E}_a(Y)
\]
for the predictor and response means. Using these means for centering, the linear prediction rule associated with \(\boldsymbol{\beta}\in\mathbb R^p\) is
\[
  \boldsymbol{x}\longmapsto \nu_a+(\boldsymbol{x}-\boldsymbol{\mu}_a)^\top\boldsymbol{\beta}.
\]
Its population squared-error risk in domain \(a\) is
\begin{equation}
\label{eq:domain-prediction-risk}
  \mathcal R_a(\boldsymbol{\beta})
  =
  \mathbb{E}_a\!\left[
    \left\{
      Y-\nu_a-(\boldsymbol{X}-\boldsymbol{\mu}_a)^\top\boldsymbol{\beta}
    \right\}^{2}
  \right].
\end{equation}

\subsection{Target-Domain PLS and Moment Mismatch}
\label{sec:target-pls-mismatch}

For target-domain prediction under covariate shift, we seek to estimate
the population PLS coefficient \(\boldsymbol{\beta}_{\target,K}^{\mathrm{PLS}}\)
and construct its associated prediction rule
\(f_{\target,K}^{\mathrm{PLS}}\) for a fixed component count \(K\geq1\).
The coefficient is defined by minimizing the prediction risk in
\eqref{eq:domain-prediction-risk} over a response-guided subspace
constructed from target-domain moments.

For \(a\in\{\source,\target\}\), define the predictor covariance
matrix and predictor--response cross-covariance vector by
\[
  \boldsymbol{\Sigma}_a
  =
  \mathbb{E}_a\!\left\{(\boldsymbol{X}-\boldsymbol{\mu}_a)(\boldsymbol{X}-\boldsymbol{\mu}_a)^\top\right\},
  \qquad
  \boldsymbol{\sigma}_a
  =
  \mathbb{E}_a\!\left\{(\boldsymbol{X}-\boldsymbol{\mu}_a)(Y-\nu_a)\right\}.
\]
Expanding the prediction risk gives
\[
  \mathcal R_a(\boldsymbol{\beta})
  =
  \mathbb{E}_a\!\left\{(Y-\nu_a)^2\right\}
  -2\boldsymbol{\beta}^\top\boldsymbol{\sigma}_a
  +\boldsymbol{\beta}^\top\boldsymbol{\Sigma}_a\boldsymbol{\beta}.
\]
The dependence of the risk on \(\boldsymbol{\beta}\) is therefore determined
by \(\boldsymbol{\Sigma}_a\) and \(\boldsymbol{\sigma}_a\).
Population PLS uses these same moments for dimension reduction
by restricting the coefficient vector to the Krylov
subspace~\citep{CookHellandSu2013}
\[
  \mathcal K_{a,K}
  =
  \operatorname{span}
  \left\{
    \boldsymbol{\sigma}_a,\,
    \boldsymbol{\Sigma}_a\boldsymbol{\sigma}_a,\,
    \ldots,\,
    \boldsymbol{\Sigma}_a^{K-1}\boldsymbol{\sigma}_a
  \right\}.
\]

For any \(\boldsymbol{v}\in\ker(\boldsymbol{\Sigma}_a)\),
\(\boldsymbol{v}^\top(\boldsymbol{X}-\boldsymbol{\mu}_a)=0\) almost surely, and hence
\(\boldsymbol{v}^\top\boldsymbol{\sigma}_a=0\).
It follows that
\(\boldsymbol{\sigma}_a\in\operatorname{range}(\boldsymbol{\Sigma}_a)\) and
\(\mathcal K_{a,K}\subseteq\operatorname{range}(\boldsymbol{\Sigma}_a)\).
Since \(\boldsymbol{\Sigma}_a\) is positive definite on its range,
\(\mathcal R_a\) is strictly convex on \(\mathcal K_{a,K}\).
The population PLS coefficient in domain \(a\) is therefore
uniquely defined by
\begin{equation}
\label{eq:population-pls-minimizer}
  \boldsymbol{\beta}_{a,K}^{\mathrm{PLS}}
  =
  \operatorname*{arg\,min}_{\beta\in\mathcal K_{a,K}}
  \mathcal R_a(\boldsymbol{\beta}).
\end{equation}
This definition remains valid when the Krylov vectors are
linearly dependent. Once the subspace stabilizes, additional
Krylov steps leave the population coefficient unchanged.
Combining this coefficient with the predictor and response
means gives the associated prediction rule
\begin{equation}
\label{eq:population-pls-prediction}
  f_{a,K}^{\mathrm{PLS}}(\boldsymbol{x})
  =
  \nu_a+(\boldsymbol{x}-\boldsymbol{\mu}_a)^\top\boldsymbol{\beta}_{a,K}^{\mathrm{PLS}}.
\end{equation}

Under covariate shift, the conditional response distribution
is shared across domains, but the moments defining the PLS
subspace and optimization criterion need not coincide.
To identify the role of the predictor distribution, let
\[
  m(\boldsymbol{x})=\mathbb{E}_{\source}(Y\mid \boldsymbol{X}=\boldsymbol{x})
\]
denote the common conditional mean, which also equals
\(\mathbb{E}_{\target}(Y\mid \boldsymbol{X}=\boldsymbol{x})\) for
\(P_{\target}^{X}\)-almost every \(\boldsymbol{x}\).
By iterated expectation,
\[
  \nu_a=\mathbb{E}_a\{m(\boldsymbol{X})\},
  \qquad
  \boldsymbol{\sigma}_a
  =
  \mathbb{E}_a\!\left[
    (\boldsymbol{X}-\boldsymbol{\mu}_a)\{m(\boldsymbol{X})-\nu_a\}
  \right].
\]
Thus, \(\boldsymbol{\sigma}_a\) depends on both the shared conditional mean
\(m\) and the predictor distribution \(P_a^X\), whereas
\(\boldsymbol{\Sigma}_a\) depends on the latter alone.
Changes in the predictor distribution can therefore alter the
Krylov subspace and the corresponding population PLS coefficient.
To quantify such differences at the population level, we define
transfer discrepancy for a general statistical functional.

\noindent\textbf{Definition (Population transfer discrepancy, PTD).}
Let \(\mathcal T:\mathcal P\to\mathbb R^q\) be a statistical
functional defined on a class of distributions \(\mathcal P\).
For source and target distributions
\(P_{\source},P_{\target}\in\mathcal P\), the population transfer
discrepancy associated with \(\mathcal T\) is
\begin{equation}
\label{eq:population-transfer-discrepancy}
  \Delta_{\mathcal T}(P_{\source},P_{\target})
  =
  \mathcal T(P_{\source})-\mathcal T(P_{\target}).
\end{equation}
For a fixed functional \(\mathcal T\), the PTD quantifies the
difference between its source- and target-domain values.
It is zero when these values coincide, even if
\(P_{\source}\neq P_{\target}\), and is nonzero otherwise.

For a fixed component count \(K\), we apply this definition to
the population PLS coefficient functional
\[
  \mathcal T_K^{\mathrm{PLS}}(P_a)
  =
  \boldsymbol{\beta}_{a,K}^{\mathrm{PLS}}.
\]
Because target responses are unavailable, the target-domain PLS
coefficient cannot be estimated directly from the target sample
alone. One possible baseline is to approximate
\(\boldsymbol{\beta}_{\target,K}^{\mathrm{PLS}}\) by its source-domain
counterpart \(\boldsymbol{\beta}_{\source,K}^{\mathrm{PLS}}\).
The resulting population approximation error is the PTD
associated with \(\mathcal T_K^{\mathrm{PLS}}\):
\begin{equation}
\label{eq:population-pls-mismatch}
  \begin{aligned}
    \Delta_K^{\mathrm{shift}}
    &=
    \Delta_{\mathcal T_K^{\mathrm{PLS}}}
      (P_{\source},P_{\target})\\
    &=
    \boldsymbol{\beta}_{\source,K}^{\mathrm{PLS}}
    -
    \boldsymbol{\beta}_{\target,K}^{\mathrm{PLS}}.
  \end{aligned}
\end{equation}

Implementing this approximation requires estimating the
source-domain population coefficient from the labeled source
sample. Let \(\widehat{\boldsymbol{\beta}}_{\source,K}^{\mathrm{PLS}}\) denote
the ordinary PLS estimator constructed from the unweighted
empirical moments of that sample.
Replacing the population proxy with this estimator introduces
source-domain estimation error in addition to the population
approximation error above.
Consequently, its error relative to the target-domain
coefficient admits the exact decomposition
\begin{equation}
\label{eq:source-pls-target-error-decomposition}
\begin{aligned}
  \widehat{\boldsymbol{\beta}}_{\source,K}^{\mathrm{PLS}}
  -\boldsymbol{\beta}_{\target,K}^{\mathrm{PLS}}
  &=
  \underbrace{
    \widehat{\boldsymbol{\beta}}_{\source,K}^{\mathrm{PLS}}
    -\boldsymbol{\beta}_{\source,K}^{\mathrm{PLS}}
  }_{\text{source-domain estimation error}}
  +
  \underbrace{
    \Delta_K^{\mathrm{shift}}
  }_{\text{population transfer discrepancy}}.
\end{aligned}
\end{equation}

By \eqref{eq:source-pls-target-error-decomposition}, if the
source-domain estimation error converges to zero in probability
in a norm \(\|\cdot\|\), then ordinary source-domain PLS is
consistent for \(\boldsymbol{\beta}_{\target,K}^{\mathrm{PLS}}\) in the same
norm if and only if \(\|\Delta_K^{\mathrm{shift}}\|\to0\) along
the same asymptotic sequence.
If the latter condition fails, consistency within the source
domain does not yield consistency for the target-domain
coefficient.

The distinction between population transfer discrepancy and
within-domain approximation error is particularly clear under
the common linear conditional mean
\[
  m(\boldsymbol{x})=\alpha_0+\boldsymbol{x}^\top\boldsymbol{\beta}_0.
\]
In either domain,
\[
  \boldsymbol{\sigma}_a=\boldsymbol{\Sigma}_a\boldsymbol{\beta}_0,
  \qquad
  \mathcal K_{a,K}
  =
  \operatorname{span}
  \left\{
    \boldsymbol{\Sigma}_a\boldsymbol{\beta}_0,\,
    \boldsymbol{\Sigma}_a^2\boldsymbol{\beta}_0,\,
    \ldots,\,
    \boldsymbol{\Sigma}_a^K\boldsymbol{\beta}_0
  \right\}.
\]
Under this model, the excess prediction risk of the population
PLS coefficient is
\[
  \mathcal R_a(\boldsymbol{\beta}_{a,K}^{\mathrm{PLS}})
  -
  \mathcal R_a(\boldsymbol{\beta}_0)
  =
  (\boldsymbol{\beta}_{a,K}^{\mathrm{PLS}}-\boldsymbol{\beta}_0)^\top
  \boldsymbol{\Sigma}_a
  (\boldsymbol{\beta}_{a,K}^{\mathrm{PLS}}-\boldsymbol{\beta}_0).
\]
Here, \(\boldsymbol{\beta}_{a,K}^{\mathrm{PLS}}-\boldsymbol{\beta}_0\) is the
finite-component approximation error within domain \(a\),
and \(\boldsymbol{\Sigma}_a\) determines its contribution to prediction risk.
The PTD \(\Delta_K^{\mathrm{shift}}\), by comparison, measures
the difference between the source- and target-domain
approximations to the same underlying coefficient \(\boldsymbol{\beta}_0\).

These observations motivate reconstructing the target-domain
moments before PLS component extraction.
Under the covariate-shift assumptions in
\eqref{eq:covariate-shift-setting}, these moments can be recovered
as importance-weighted source-domain expectations, thereby
identifying the target-domain PLS functional.
The next subsection establishes the corresponding identities
for the target-domain means, predictor covariance, and
predictor--response cross-covariance, providing the population
foundation for IWPLS.

\subsection{Population Formulation of IWPLS via Importance Weighting}

To address the population transfer discrepancy in
\eqref{eq:population-pls-mismatch}, we reconstruct the
target-domain moments before forming the PLS components.
Under \eqref{eq:covariate-shift-setting}, this reconstruction
is achieved by reweighting the source joint distribution
using the target-to-source covariate density ratio.

The absolute-continuity condition
\(P_{\target}^{X}\ll P_{\source}^{X}\) ensures the existence
of the Radon--Nikodym derivative
\begin{equation}
\label{eq:setup-density-ratio}
  w_0(\boldsymbol{x})
  =
  \frac{dP_{\target}^{X}}{dP_{\source}^{X}}(\boldsymbol{x}).
\end{equation}
This ratio is unique \(P_{\source}^{X}\)-almost everywhere
and satisfies
\(\mathbb{E}_{\source}\{w_0(\boldsymbol{X})\}=1\).
If the predictor marginals admit densities
\(p_{\source}\) and \(p_{\target}\) with respect to a common
dominating measure, then
\[
  w_0(\boldsymbol{x})=\frac{p_{\target}(\boldsymbol{x})}{p_{\source}(\boldsymbol{x})}
\]
for \(P_{\source}^{X}\)-almost every \(\boldsymbol{x}\) with
\(p_{\source}(\boldsymbol{x})>0\).

Define the reweighted source joint probability measure
\(P_{\source}^{w_0}\) by
\[
  dP_{\source}^{w_0}(\boldsymbol{x},y)
  =
  w_0(\boldsymbol{x})\,dP_{\source}(\boldsymbol{x},y).
\]
Its predictor marginal is \(P_{\target}^{X}\).
Since weighting by a function of \(\boldsymbol{X}\) preserves the
conditional response law, the conditional invariance in
\eqref{eq:covariate-shift-setting} implies
\[
  P_{\source}^{w_0}=P_{\target}.
\]
Consequently, for any measurable function \(g\) satisfying
\(\mathbb{E}_{\target}|g(\boldsymbol{X},Y)|<\infty\), the following
change-of-measure identity holds~\citep{Shimodaira2000CovariateShift,SugiyamaKrauledatMuller2007}:
\begin{equation}
\label{eq:setup-iw-identity}
  \mathbb{E}_{\target}\{g(\boldsymbol{X},Y)\}
  =
  \mathbb{E}_{\source}\{w_0(\boldsymbol{X})g(\boldsymbol{X},Y)\}.
\end{equation} 
Applied componentwise, this identity also covers
vector- and matrix-valued moments.

In particular, the target-domain means satisfy
\[
  \boldsymbol{\mu}_{\target}
  =
  \mathbb{E}_{\source}\{w_0(\boldsymbol{X})\boldsymbol{X}\},
  \qquad
  \nu_{\target}
  =
  \mathbb{E}_{\source}\{w_0(\boldsymbol{X})Y\},
\]
and the covariance moments satisfy
\begin{equation}
\label{eq:setup-iw-moments}
  \begin{aligned}
    \boldsymbol{\Sigma}_{\target}
    &=
    \mathbb{E}_{\source}\!\left[
      w_0(\boldsymbol{X})(\boldsymbol{X}-\boldsymbol{\mu}_{\target})(\boldsymbol{X}-\boldsymbol{\mu}_{\target})^\top
    \right],\\
    \boldsymbol{\sigma}_{\target}
    &=
    \mathbb{E}_{\source}\!\left[
      w_0(\boldsymbol{X})(\boldsymbol{X}-\boldsymbol{\mu}_{\target})(Y-\nu_{\target})
    \right].
  \end{aligned}
\end{equation}
The target predictor marginal directly identifies
\(\boldsymbol{\mu}_{\target}\) and \(\boldsymbol{\Sigma}_{\target}\).
The weighted source representation additionally identifies
the response-dependent moments
\(\nu_{\target}\) and \(\boldsymbol{\sigma}_{\target}\).
Together, these moments determine the target Krylov subspace,
its unique risk-minimizing coefficient, and the associated
prediction rule.

For fixed \(K\), define the population IWPLS coefficient
using the true density ratio by applying the PLS functional
to \(P_{\source}^{w_0}\).
Since \(P_{\source}^{w_0}=P_{\target}\),
\begin{equation}
\label{eq:setup-weighted-pls-identification}
  \begin{aligned}
    \boldsymbol{\beta}_K^{\mathrm{IWPLS}}
    &:=
    \mathcal T_K^{\mathrm{PLS}}(P_{\source}^{w_0})
    =
    \boldsymbol{\beta}_{\target,K}^{\mathrm{PLS}},\\
    \Delta_{\mathcal T_K^{\mathrm{PLS}}}
      (P_{\source}^{w_0},P_{\target})
    &=0.
  \end{aligned}
\end{equation}
The corresponding population prediction rule is
\begin{equation}
\label{eq:setup-population-iwpls-prediction}
  f_K^{\mathrm{IWPLS}}(\boldsymbol{x})
  :=
  \nu_{\target}
  +(\boldsymbol{x}-\boldsymbol{\mu}_{\target})^\top\boldsymbol{\beta}_K^{\mathrm{IWPLS}}
  =
  f_{\target,K}^{\mathrm{PLS}}(\boldsymbol{x}).
\end{equation}
Thus, the target-domain PLS coefficient and prediction rule
are identifiable from
\((P_{\source},P_{\target}^{X})\), without target responses.

Applying \eqref{eq:setup-iw-identity} to the squared-error
loss gives the equivalent weighted-risk formulation
\begin{equation}
\label{eq:setup-population-iwpls-objective}
  \boldsymbol{\beta}_K^{\mathrm{IWPLS}}
  =
  \operatorname*{arg\,min}_{\beta\in\mathcal K_{\target,K}}
  \mathbb{E}_{\source}\!\left[
    w_0(\boldsymbol{X})
    \{Y-\nu_{\target}-(\boldsymbol{X}-\boldsymbol{\mu}_{\target})^\top\boldsymbol{\beta}\}^{2}
  \right].
\end{equation}
Both the component space and the regression criterion
are therefore determined by the same reweighted source
distribution.
If components are instead extracted from unweighted source
moments, weighting only the target-centered loss minimizes
\(\mathcal R_{\target}(\boldsymbol{\beta})\) over
\(\mathcal K_{\source,K}\).
This subspace need not coincide with
\(\mathcal K_{\target,K}\), so weighting the regression
criterion alone need not recover
\(\boldsymbol{\beta}_{\target,K}^{\mathrm{PLS}}\).

\noindent\textit{Remark (Population identification and prediction risk).}
At a common component count \(K\), the relevant target-risk
comparison is between
\[
  \mathbb{E}_{\target}\!\left[
    \{Y-f_K^{\mathrm{IWPLS}}(\boldsymbol{X})\}^{2}
  \right]
  \quad\text{and}\quad
  \mathbb{E}_{\target}\!\left[
    \{Y-f_{\source,K}^{\mathrm{PLS}}(\boldsymbol{X})\}^{2}
  \right].
\]
With the true density ratio, population IWPLS recovers the
target-domain PLS coefficient and its associated prediction rule.
This removes the population coefficient approximation error
incurred by using the source-domain coefficient as a proxy.
The resulting rule minimizes target risk over target-centered
linear prediction rules with coefficients in
\(\mathcal K_{\target,K}\).

Ordinary source-domain PLS uses source-domain centering and
the source Krylov subspace \(\mathcal K_{\source,K}\), so its
prediction rule need not belong to this class.
Recovering the target functional therefore addresses the
population coefficient mismatch, while the predictive performance
of IWPLS at a fixed component count \(K\) also depends on the
quality of the target-domain finite-component approximation.

To implement this population construction, Section~3 develops
empirical IWPLS estimators based on a normalized weighted
empirical measure of the labeled source sample.
The same measure is used to construct the empirical moments,
component space, and regression criterion.
Using the true density ratio \(w_0\) yields the oracle estimators
\(\widetilde{\boldsymbol{\beta}}_K^{\mathrm{IWPLS}}\) and
\(\widetilde f_K^{\mathrm{IWPLS}}\).
Replacing \(w_0\) with an estimated density ratio yields
the estimated-weight counterparts
\(\widehat{\boldsymbol{\beta}}_K^{\mathrm{IWPLS}}\) and
\(\widehat f_K^{\mathrm{IWPLS}}\).
Density-ratio estimation uses only source and unlabeled target
covariates, while source responses enter the weighted response
mean, predictor--response cross-covariance, and regression criterion.

In this empirical construction, sampling variability perturbs
the moments defining the PLS estimator, and estimating \(w_0\)
introduces an additional source of error.
Section~4 establishes consistency for fixed \(K\) under its
model assumptions and conditions on weighted empirical moments,
density-ratio estimation error, and Krylov stability.

\section{Importance-Weighted Partial Least Squares}
\label{sec:iwpls-methodology}

For a fixed component count \(K\geq1\), we estimate the
target-domain coefficient \(\boldsymbol{\beta}_{\target,K}^{\mathrm{PLS}}\)
and prediction rule \(f_{\target,K}^{\mathrm{PLS}}\) by replacing
the population moments in Section~2.3 with self-normalized
importance-weighted source sample moments.
We first give the construction for the true density ratio
and then for an estimated ratio.
Within each construction, the same normalized weights are used
for centering, component extraction, and coefficient estimation.

\subsection{IWPLS with Oracle Importance Weights}
\label{sec:method-oracle}

Suppose that the true density ratio \(w_0\) in
\eqref{eq:setup-density-ratio} is known.
For a source sample with positive total oracle weight, define
\begin{equation}
\label{eq:method-oracle-weights}
  \widetilde\omega_i
  =
  \frac{w_0(\boldsymbol{X}_i^{\source})}
       {\sum_{\ell=1}^{n_{\source}}w_0(\boldsymbol{X}_\ell^{\source})},
  \qquad i=1,\ldots,n_{\source}.
\end{equation}
The normalized oracle weights define the weighted empirical
probability measure
\[
  \widetilde P_{\source}^{w_0}
  =
  \sum_{i=1}^{n_{\source}}
    \widetilde\omega_i
    \delta_{(X_i^{\source},Y_i^{\source})},
\]
where \(\delta_{(x,y)}\) denotes the point mass at \((\boldsymbol{x},y)\).
Applying the population PLS construction to this measure
gives the oracle IWPLS estimator.
Its empirical moments therefore provide a common basis
for centering, component extraction, and coefficient fitting.
In particular, the predictor and response means under this
measure are
\begin{equation}
\label{eq:method-oracle-means}
  \widetilde{\boldsymbol{\mu}}
  =
  \sum_{i=1}^{n_{\source}}
    \widetilde\omega_i\boldsymbol{X}_i^{\source},
  \qquad
  \widetilde\nu
  =
  \sum_{i=1}^{n_{\source}}
    \widetilde\omega_iY_i^{\source}.
\end{equation}

These means define the weighted predictor covariance matrix
and predictor--response cross-covariance vector
\begin{equation}
\label{eq:method-oracle-moments}
\begin{aligned}
  \widetilde{\boldsymbol{\Sigma}}
  &=
  \sum_{i=1}^{n_{\source}}\widetilde\omega_i
    (\boldsymbol{X}_i^{\source}-\widetilde{\boldsymbol{\mu}})
    (\boldsymbol{X}_i^{\source}-\widetilde{\boldsymbol{\mu}})^\top,\\
  \widetilde{\boldsymbol{\sigma}}
  &=
  \sum_{i=1}^{n_{\source}}\widetilde\omega_i
    (\boldsymbol{X}_i^{\source}-\widetilde{\boldsymbol{\mu}})
    (Y_i^{\source}-\widetilde\nu).
\end{aligned}
\end{equation}
These are the self-normalized empirical counterparts of the
target moments in \eqref{eq:setup-iw-moments}.

Using the same weights and centering, define
\begin{equation}
\label{eq:method-oracle-risk}
\begin{aligned}
  \widetilde{\mathcal R}(\boldsymbol{\beta})
  &=
  \sum_{i=1}^{n_{\source}}\widetilde\omega_i
    \left\{
      Y_i^{\source}-\widetilde\nu
      -(\boldsymbol{X}_i^{\source}-\widetilde{\boldsymbol{\mu}})^\top\boldsymbol{\beta}
    \right\}^{2}\\
  &=
  \sum_{i=1}^{n_{\source}}\widetilde\omega_i
    (Y_i^{\source}-\widetilde\nu)^2
  -2\boldsymbol{\beta}^\top\widetilde{\boldsymbol{\sigma}}
  +\boldsymbol{\beta}^\top\widetilde{\boldsymbol{\Sigma}}\boldsymbol{\beta}.
\end{aligned}
\end{equation}
The same moment pair generates the empirical Krylov matrix
and subspace
\[
  \widetilde {\mathbf{G}}_K
  =
  \left(
    \widetilde{\boldsymbol{\sigma}},\,
    \widetilde{\boldsymbol{\Sigma}}\widetilde{\boldsymbol{\sigma}},\,
    \ldots,\,
    \widetilde{\boldsymbol{\Sigma}}^{K-1}\widetilde{\boldsymbol{\sigma}}
  \right),
  \qquad
  \widetilde{\mathcal K}_K
  =
  \operatorname{col}(\widetilde {\mathbf{G}}_K).
\]
As in Section~\ref{sec:target-pls-mismatch},
\(\widetilde{\boldsymbol{\sigma}}\in\operatorname{range}(\widetilde{\boldsymbol{\Sigma}})\),
so \(\widetilde{\mathcal K}_K\) is contained in this range.
The restriction of \(\widetilde{\boldsymbol{\Sigma}}\) to this subspace is
positive definite, and the quadratic criterion has a unique
minimizer there.
The oracle IWPLS coefficient is therefore
\begin{equation}
\label{eq:method-oracle-objective}
  \widetilde{\boldsymbol{\beta}}_K^{\mathrm{IWPLS}}
  =
  \operatorname*{arg\,min}_{\beta\in\widetilde{\mathcal K}_K}
    \widetilde{\mathcal R}(\boldsymbol{\beta}).
\end{equation}
The minimizer remains unique when \(\widetilde{\boldsymbol{\Sigma}}\) is
singular or the Krylov columns are linearly dependent.

Let
\[
  \widetilde{\boldsymbol{\Theta}}_K
  =
  \widetilde {\mathbf{G}}_K^\top
  \widetilde{\boldsymbol{\Sigma}}
  \widetilde {\mathbf{G}}_K.
\]
When \(\widetilde {\mathbf{G}}_K\) has full column rank,
\(\widetilde{\boldsymbol{\Theta}}_K\) is positive definite, and the restricted
normal equations yield
\begin{equation}
\label{eq:method-oracle-fit}
  \widetilde{\boldsymbol{\beta}}_K^{\mathrm{IWPLS}}
  =
  \widetilde {\mathbf{G}}_K
  \widetilde{\boldsymbol{\Theta}}_K^{-1}
  \widetilde {\mathbf{G}}_K^\top
  \widetilde{\boldsymbol{\sigma}}.
\end{equation}
If the Krylov columns are dependent and
\(r_K=\dim(\widetilde{\mathcal K}_K)>0\), any basis matrix
\(\widetilde {\mathbf{B}}_K\in\mathbb R^{p\times r_K}\) gives the
equivalent representation
\[
  \widetilde{\boldsymbol{\beta}}_K^{\mathrm{IWPLS}}
  =
  \widetilde {\mathbf{B}}_K
  \left(
    \widetilde {\mathbf{B}}_K^\top
    \widetilde{\boldsymbol{\Sigma}}
    \widetilde {\mathbf{B}}_K
  \right)^{-1}
  \widetilde {\mathbf{B}}_K^\top
  \widetilde{\boldsymbol{\sigma}}.
\]
If the subspace is \(\{0\}\), the coefficient is zero.
Once the Krylov subspace stabilizes, increasing \(K\) leaves
the coefficient unchanged.
The associated prediction rule is
\begin{equation}
\label{eq:method-oracle-prediction}
  \widetilde f_K^{\mathrm{IWPLS}}(\boldsymbol{x})
  =
  \widetilde\nu
  +(\boldsymbol{x}-\widetilde{\boldsymbol{\mu}})^\top
    \widetilde{\boldsymbol{\beta}}_K^{\mathrm{IWPLS}}.
\end{equation}

For \(K=1\) and \(\widetilde{\boldsymbol{\sigma}}\neq0\),
\begin{equation}
\label{eq:method-oracle-one-component}
  \widetilde{\boldsymbol{\beta}}_1^{\mathrm{IWPLS}}
  =
  \frac{
    \widetilde{\boldsymbol{\sigma}}^\top\widetilde{\boldsymbol{\sigma}}
  }{
    \widetilde{\boldsymbol{\sigma}}^\top
    \widetilde{\boldsymbol{\Sigma}}
    \widetilde{\boldsymbol{\sigma}}
  }
  \widetilde{\boldsymbol{\sigma}}.
\end{equation}
The covariance-range property ensures a positive denominator.
Section~4 uses this representation for the one-component
consistency and projected asymptotic analysis.
When \(w_0=1\) \(P_{\source}^{X}\)-almost everywhere,
\(\widetilde\omega_i=1/n_{\source}\), and oracle IWPLS
coincides with ordinary source-domain PLS under the same
preprocessing conventions and component count.
Using the true density ratio, the oracle estimator provides
a benchmark for assessing the additional error introduced
by weight estimation.

\subsection{IWPLS with Estimated Importance Weights}
\label{sec:method-estimated}

In practice, \(w_0\) is unknown.
Let \(\widehat w:\mathbb R^p\to[0,\infty)\) be a nonnegative
estimate of \(w_0\) obtained from source and unlabeled
target covariates.
The construction below accommodates different density-ratio
fitting procedures.
The procedure used in our numerical studies is
described in Supplementary Section~A.1.
When the total estimated source weight is finite and positive,
define
\begin{equation}
\label{eq:method-estimated-weights}
  \widehat\omega_i
  =
  \frac{\widehat w(\boldsymbol{X}_i^{\source})}
       {\sum_{\ell=1}^{n_{\source}}
         \widehat w(\boldsymbol{X}_\ell^{\source})},
  \qquad i=1,\ldots,n_{\source}.
\end{equation}
Replacing \(\widetilde\omega_i\) by \(\widehat\omega_i\) in
\eqref{eq:method-oracle-means} and
\eqref{eq:method-oracle-moments} defines
\(\widehat{\boldsymbol{\mu}},\widehat\nu,\widehat{\boldsymbol{\Sigma}},\widehat{\boldsymbol{\sigma}}\).
All four quantities are moments of the normalized weighted
empirical measure
\[
  \widehat P_{\source}^{\widehat w}
  =
  \sum_{i=1}^{n_{\source}}
    \widehat\omega_i
    \delta_{(X_i^{\source},Y_i^{\source})}.
\]
By \eqref{eq:method-estimated-weights}, a common positive
rescaling of \(\widehat w\) preserves the normalized weights
and therefore leaves this empirical measure and all four
moments unchanged.

For a fixed component count \(K\geq1\), these moments determine
the empirical Krylov matrix and subspace
\[
  \widehat {\mathbf{G}}_K
  =
  \left(
    \widehat{\boldsymbol{\sigma}},\,
    \widehat{\boldsymbol{\Sigma}}\widehat{\boldsymbol{\sigma}},\,
    \ldots,\,
    \widehat{\boldsymbol{\Sigma}}^{K-1}\widehat{\boldsymbol{\sigma}}
  \right),
  \qquad
  \widehat{\mathcal K}_K
  =
  \operatorname{col}(\widehat {\mathbf{G}}_K).
\]
The coefficient is fitted over this subspace using the
weighted empirical risk
\[
  \widehat{\mathcal R}(\boldsymbol{\beta})
  =
  \sum_{i=1}^{n_{\source}}\widehat\omega_i
    \left\{
      Y_i^{\source}-\widehat\nu
      -(\boldsymbol{X}_i^{\source}-\widehat{\boldsymbol{\mu}})^\top\boldsymbol{\beta}
    \right\}^{2}.
\]
The estimated-weight IWPLS coefficient is
\begin{equation}
\label{eq:method-estimated-objective}
  \widehat{\boldsymbol{\beta}}_K^{\mathrm{IWPLS}}
  =
  \operatorname*{arg\,min}_{\beta\in\widehat{\mathcal K}_K}
    \widehat{\mathcal R}(\boldsymbol{\beta}),
\end{equation}
with associated prediction rule
\begin{equation}
\label{eq:method-feasible-prediction}
  \widehat f_K^{\mathrm{IWPLS}}(\boldsymbol{x})
  =
  \widehat\nu
  +(\boldsymbol{x}-\widehat{\boldsymbol{\mu}})^\top
    \widehat{\boldsymbol{\beta}}_K^{\mathrm{IWPLS}}.
\end{equation}
Uniqueness follows from the same covariance-range argument
as in the oracle construction.

Let
\[
  \widehat{\boldsymbol{\Theta}}_K
  =
  \widehat {\mathbf{G}}_K^\top
  \widehat{\boldsymbol{\Sigma}}
  \widehat {\mathbf{G}}_K.
\]
When \(\widehat {\mathbf{G}}_K\) has full column rank,
\(\widehat{\boldsymbol{\Theta}}_K\) is positive definite, and
\begin{equation}
\label{eq:method-feasible-fit}
  \widehat{\boldsymbol{\beta}}_K^{\mathrm{IWPLS}}
  =
  \widehat {\mathbf{G}}_K
  \widehat{\boldsymbol{\Theta}}_K^{-1}
  \widehat {\mathbf{G}}_K^\top
  \widehat{\boldsymbol{\sigma}}.
\end{equation}
If the generating columns are dependent, the coefficient
is evaluated using a basis of the actual Krylov subspace,
as in the oracle construction.
If that subspace is \(\{0\}\), the coefficient is zero.
For \(K=1\) and \(\widehat{\boldsymbol{\sigma}}\neq0\),
\begin{equation}
\label{eq:method-one-component}
  \widehat{\boldsymbol{\beta}}_1^{\mathrm{IWPLS}}
  =
  \frac{
    \widehat{\boldsymbol{\sigma}}^\top\widehat{\boldsymbol{\sigma}}
  }{
    \widehat{\boldsymbol{\sigma}}^\top
    \widehat{\boldsymbol{\Sigma}}
    \widehat{\boldsymbol{\sigma}}
  }
  \widehat{\boldsymbol{\sigma}}.
\end{equation}

For high-dimensional computation, we express both the weighted
Krylov construction and the regression fit in terms of
transformed source observations.
Centering the source observations by
\((\widehat{\boldsymbol{\mu}},\widehat\nu)\) and scaling each centered
predictor--response pair by \(\sqrt{\widehat\omega_i}\)
gives a design matrix \(\mathbf{X}^*\) and response vector \(\boldsymbol{Y}^*\)
satisfying
\[
  (\mathbf{X}^*)^\top \mathbf{X}^*=\widehat{\boldsymbol{\Sigma}},
  \qquad
  (\mathbf{X}^*)^\top \boldsymbol{Y}^*=\widehat{\boldsymbol{\sigma}},
  \qquad
  \|\boldsymbol{Y}^*-\mathbf{X}^*\boldsymbol{\beta}\|_2^2=\widehat{\mathcal R}(\boldsymbol{\beta}).
\]
These identities imply that PLS for a univariate response
(PLS1), applied without additional centering, scaling, or
intercept fitting, constructs
\(\widehat{\mathcal K}_K\) implicitly and solves the restricted
minimization in \eqref{eq:method-estimated-objective}.
The IWPLS fit can therefore be computed without explicitly
forming \(\widehat{\boldsymbol{\Sigma}}\) or its powers.

Algorithm~\ref{alg:iwpls-fixed-k} implements the IWPLS fit
for a given component count \(K\) by normalizing the supplied
importance weights, centering and transforming the source
observations, and applying PLS1.
We treat \(K\) as a tuning parameter and use importance-weighted
cross-validation
(IWCV)~\citep{SugiyamaKrauledatMuller2007}
to select \(\widehat K\) from a finite candidate set
\(\mathcal A_0\subset\{1,2,\ldots\}\) for target-domain prediction.
The full-source fit at \(K=\widehat K\) yields
\(\widehat{\boldsymbol{\beta}}_{\widehat K}^{\mathrm{IWPLS}}\) and
\(\widehat f_{\widehat K}^{\mathrm{IWPLS}}\), with original-scale
intercept
\(\widehat\nu-\widehat{\boldsymbol{\mu}}^\top
  \widehat{\boldsymbol{\beta}}_{\widehat K}^{\mathrm{IWPLS}}\).
Model fitting and component selection use labeled source
observations and unlabeled target adaptation covariates,
while predictive performance is assessed on an independent
target test sample.
\begin{algorithm}[htbp]
\caption{Fixed-\(K\) IWPLS fitting}
\label{alg:iwpls-fixed-k}
\small
\noindent\textbf{Input:} Labeled source sample
\(\mathcal D_{\source}\), nonnegative raw weights
\(\{\widehat w(\boldsymbol{X}_i^{\source})\}_{i=1}^{n_{\source}}\),
and a fixed component count \(K\geq1\).
\par\smallskip
\noindent\textbf{Output:} Coefficient
\(\widehat{\boldsymbol{\beta}}_K^{\mathrm{IWPLS}}\) and prediction rule
\(\widehat f_K^{\mathrm{IWPLS}}\), or a fitting failure.
\begin{enumerate}
  \setlength{\itemsep}{3pt}
  \setlength{\parskip}{0pt}

  \item \textit{Normalize and center.}
  Check that the total raw weight is finite and positive.
  Otherwise, report fitting failure and stop.
  Compute \(\widehat\omega_i\) by
  \eqref{eq:method-estimated-weights} and the weighted
  means \(\widehat{\boldsymbol{\mu}}\) and \(\widehat\nu\).

  \item \textit{Transform the observations.}
  For \(i=1,\ldots,n_{\source}\), set
  \begin{equation}
  \label{eq:method-transformed-data}
    \boldsymbol{X}_i^*
    =
    \sqrt{\widehat\omega_i}
      (\boldsymbol{X}_i^{\source}-\widehat{\boldsymbol{\mu}}),
    \qquad
    Y_i^*
    =
    \sqrt{\widehat\omega_i}
      (Y_i^{\source}-\widehat\nu).
  \end{equation}
  Form \(\mathbf{X}^*\in\mathbb R^{n_{\source}\times p}\) with rows
  \((\boldsymbol{X}_i^*)^\top\), and let
  \(\boldsymbol{Y}^*=(Y_1^*,\ldots,Y_{n_{\source}}^*)^\top\).

  \item \textit{Fit the coefficient.}
  If \((\mathbf{X}^*)^\top \boldsymbol{Y}^*=0\), set
  \(\widehat{\boldsymbol{\beta}}_K^{\mathrm{IWPLS}}=0\).
  Otherwise, run PLS1 for up to \(K\) components with
  additional centering, scaling, and intercept fitting disabled.
  If the Krylov subspace stabilizes before \(K\), retain
  the coefficient from the last nondegenerate fit.
  If numerical breakdown occurs or the coefficient is
  nonfinite, report fitting failure and stop.

  \item \textit{Return the predictor.}
  Return \(\widehat{\boldsymbol{\beta}}_K^{\mathrm{IWPLS}}\) and
  \(\widehat f_K^{\mathrm{IWPLS}}(\boldsymbol{x})
    =\widehat\nu+(\boldsymbol{x}-\widehat{\boldsymbol{\mu}})^\top
      \widehat{\boldsymbol{\beta}}_K^{\mathrm{IWPLS}}\).
\end{enumerate}
\end{algorithm}

For the theoretical analysis, we consider a fixed component
count \(K\).
The oracle estimator computed from the same labeled source
sample provides a reference for separating the effect of
weight estimation from sampling error.
Specifically, the error relative to the target-domain
population PLS coefficient admits the exact decomposition
\begin{equation}
\label{eq:method-error-decomposition}
\begin{aligned}
  \widehat{\boldsymbol{\beta}}_K^{\mathrm{IWPLS}}
  -\boldsymbol{\beta}_{\target,K}^{\mathrm{PLS}}
  &=
  \underbrace{
    \widehat{\boldsymbol{\beta}}_K^{\mathrm{IWPLS}}
    -\widetilde{\boldsymbol{\beta}}_K^{\mathrm{IWPLS}}
  }_{\text{weight-estimation effect}}\\
  &\quad+
  \underbrace{
    \widetilde{\boldsymbol{\beta}}_K^{\mathrm{IWPLS}}
    -\boldsymbol{\beta}_{\target,K}^{\mathrm{PLS}}
  }_{\text{oracle sampling error}}.
\end{aligned}
\end{equation}
The first term captures the perturbation induced by weight
estimation through normalization, centering, weighted moments,
component extraction, and coefficient fitting.
The second is the sampling error of oracle IWPLS.
Section~4 analyzes both terms for fixed \(K\) under conditions
on weighted empirical moments, weight-estimation error,
and Krylov stability.

\section{Theoretical Results}

Building on the error decomposition in
\eqref{eq:method-error-decomposition},
we establish consistency in the target covariance norm
for oracle and estimated-weight IWPLS with fixed
component counts below both domain-specific envelope
dimensions. We also derive prediction-error bounds
for the corresponding prediction rules relative to
the target population PLS rule. For the one-component
case, we further establish projected asymptotic
normality under oracle weights and, with additional
conditions on density-ratio estimation, under
estimated weights.

Our analysis combines perturbation bounds for
self-normalized weighted moments with stability
arguments for the normalized Krylov system.
The consistency results cover fixed component counts,
including $K=1$ (Supplementary Sections~E and~H).
For $K=1$, influence-function expansions yield the
oracle and estimated-weight distributional results
(Supplementary Sections~F and~G).

\subsection{Model and Assumptions}
\label{sec:theory-moments}

For the theoretical analysis of IWPLS, assume
\begin{equation}
\label{eq:theory-model}
Y^a=\alpha_0+
(\boldsymbol X^a)^\top\boldsymbol\beta_0
+\varepsilon^a,
\qquad
\boldsymbol X^a\sim
N_p(\boldsymbol\mu_a,\boldsymbol\Sigma_a),
\quad a\in\{\source,\target\},
\end{equation}
where $\alpha_0\in\mathbb R$ and
$\boldsymbol\beta_0\in\mathbb R^p$ are
common to both domains, while
$\boldsymbol\mu_a$ and
$\boldsymbol\Sigma_a\succ0$ may differ.
The errors are independent of the
covariates and share a common distribution
with mean zero and variance $\tau^2>0$.
Consequently,
\[
\nu_a=\alpha_0+
\boldsymbol\mu_a^\top\boldsymbol\beta_0,
\qquad
\boldsymbol\sigma_a=
\boldsymbol\Sigma_a\boldsymbol\beta_0.
\]

The labeled source and unlabeled target
samples are independent, with i.i.d.
observations within each domain and
common size $n_{\source}=n_{\target}=n$.
We consider $n\to\infty$ and
$p=p_n\to\infty$, allowing $p>n$,
dense regression coefficients, and
strongly correlated predictors.
Distributions and parameters may vary
along the asymptotic sequence.
All constants in the stated bounds are
uniform in $n$ and $p$, and
$a_n\asymp b_n$ means that their ratio
is bounded above and below by positive
constants.

For $a\in\{\source,\target\}$, let
\begin{equation}
\label{eq:theory-envelope}
\mathcal E_a=
\mathcal E_{\boldsymbol\Sigma_a}
\bigl(\operatorname{span}
(\boldsymbol\beta_0)\bigr),
\qquad
d_a=\dim(\mathcal E_a),
\end{equation}
where $\mathcal E_a$ is the smallest
reducing subspace of
$\boldsymbol\Sigma_a$ containing
$\boldsymbol\beta_0$~
\citep{CookHellandSu2013,CookForzani2019}.
The envelope dimensions may differ
across domains.

For weighted-moment analysis, define
\[
\boldsymbol Z=
\boldsymbol X-\boldsymbol\mu_{\target},
\qquad
U=Y-\nu_{\target},
\qquad
\mathbb P_{\source,n}g=
\frac1n\sum_{i=1}^n
g(\boldsymbol X_i^{\source},
Y_i^{\source}).
\]
Equation~\eqref{eq:setup-iw-moments}
identifies the target moments as
weighted source moments of
$(\boldsymbol Z,U)$.
The estimators use the weighted
sample means for centering, as defined
in Section~\ref{sec:iwpls-methodology}.
Coefficient error is measured in
\[
\|\boldsymbol v\|_{\boldsymbol\Sigma_{\target}}
=
(\boldsymbol v^\top
\boldsymbol\Sigma_{\target}
\boldsymbol v)^{1/2}.
\]

\begin{assumption}[Oracle weighted moments]
\label{assumption:oracle-moments}
The target moments satisfy
\[
\|\boldsymbol\sigma_{\target}\|_2^2
=O(p),
\qquad
\operatorname{tr}
(\boldsymbol\Sigma_{\target}^2)
=O(p^2).
\]
The weighted source moments satisfy
\[
\mathbb{E}_{\source}w_0^2=O(1),
\qquad
\mathbb{E}_{\source}(w_0^2U^2)=O(1),
\]
and
\[
\begin{aligned}
\mathbb{E}_{\source}
(w_0^2\|\boldsymbol Z\|_2^4)
&=O(p^2),\\
\mathbb{E}_{\source}
(w_0^2\|\boldsymbol Z\|_2^2U^2)
&=O(p).
\end{aligned}
\]
\end{assumption}

These conditions control fluctuations
of the self-normalized weighted means,
covariance, and cross-covariance.
They allow dense coefficients and
growing covariance eigenvalues.

Let $\widehat w$ be a nonnegative
density-ratio estimator based on source
and unlabeled target covariates. On
\[
\mathcal E_{w,n}
=\{0<P_{\source}^X\widehat w<\infty\},
\]
define
\[
\omega_n=
\frac{\widehat w}{P_{\source}^X\widehat w},
\qquad
h_n=\omega_n-w_0.
\]
The population normalization is used
only for analysis; it does not alter
the self-normalized empirical weights.

\begin{assumption}[Empirical weight error]
\label{assumption:estimated-weight}
Assume $\Pr(\mathcal E_{w,n})\to1$.
On $\mathcal E_{w,n}^c$, set
$\omega_n=w_0$ and $h_n=0$ for
theoretical analysis.
Suppose a deterministic sequence
$r_w=r_w(n,p)\to0$ satisfies
\[
\|h_n\|_{n,2}
:=
\{\mathbb P_{\source,n}h_n^2\}^{1/2}
=O_{\mathbb P}(r_w).
\]
In addition, the unweighted source
moments satisfy
\[
\begin{aligned}
\mathbb{E}_{\source}\|\boldsymbol Z\|_2^4
&=O(p^2),\qquad
\mathbb{E}_{\source}U^2=O(1),\qquad
\mathbb{E}_{\source}
(\|\boldsymbol Z\|_2^2U^2)
&=O(p).
\end{aligned}
\]
\end{assumption}

The empirical condition permits
reuse of the source covariates in
density-ratio estimation; a population
$L_2(P_{\source}^X)$ rate alone does
not generally imply this bound.
Supplementary Section~D establishes
the resulting weighted-moment bounds,
while Supplementary Section~I.2
verifies the condition for correctly
specified fixed-basis uLSIF.

\subsection{One-Component IWPLS: Consistency and Asymptotic Distribution}
\label{sec:theory-one}

\subsubsection{One-Component Consistency}

For \(K=1<\min\{d_{\source},d_{\target}\}\), define
\[
 q_{\target}=\boldsymbol{\sigma}_{\target}^\top\boldsymbol{\sigma}_{\target},\qquad
 m_{\target}=\boldsymbol{\sigma}_{\target}^\top\boldsymbol{\Sigma}_{\target}\boldsymbol{\sigma}_{\target}.
\]
For \(q_{\target}>0\), the target population coefficient is
\[
 \boldsymbol{\beta}_{\target,1}^{\mathrm{PLS}}
 =\frac{q_{\target}}{m_{\target}}\boldsymbol{\sigma}_{\target}.
\]
The oracle and estimated-weight coefficients are given in
\eqref{eq:method-oracle-one-component} and
\eqref{eq:method-one-component}, respectively.
This scalar ratio underlies both the consistency argument and the
distributional expansions below.

\begin{assumption}[One-component scale]
\label{assumption:one-scale}
As \(n,p\to\infty\),
\[
 q_{\target}\asymp p,\qquad m_{\target}\asymp p^2.
\]
\end{assumption}

The scale condition gives \(q_{\target}/m_{\target}\asymp p^{-1}\),
which sets the multiplier scale in the ratio expansion.

\begin{theorem}[One-component consistency]
\label{thm:one-consistency}
Under the model in \eqref{eq:theory-model} and
Assumptions~\ref{assumption:oracle-moments} and~\ref{assumption:one-scale},
\[
 \|\widetilde{\boldsymbol{\beta}}_1^{\mathrm{IWPLS}}-\boldsymbol{\beta}_{\target,1}^{\mathrm{PLS}}\|_{\boldsymbol{\Sigma}_{\target}}
 =O_{\mathbb P}(n^{-1/2}).
\]
If Assumption~\ref{assumption:estimated-weight} also holds, then
\[
 \|\widehat{\boldsymbol{\beta}}_1^{\mathrm{IWPLS}}-\boldsymbol{\beta}_{\target,1}^{\mathrm{PLS}}\|_{\boldsymbol{\Sigma}_{\target}}
 =O_{\mathbb P}(n^{-1/2}+r_w).
\]
Under their respective assumptions, the scalar denominators in the
oracle and estimated-weight coefficient representations are positive
with probability tending to one.
\end{theorem}

The two rates separate oracle sampling error from the additional effect
of weight estimation. The scale condition is compatible with dense
coefficients and signals spanning multiple covariance eigenspaces.
Section~\ref{sec:theory-prediction} gives the prediction consequences.

\subsubsection{Oracle Asymptotic Distribution}

For \(K=1\), we study first-order fluctuations of the oracle coefficient
error relative to \(\boldsymbol{\beta}_{\target,1}^{\mathrm{PLS}}\) along deterministic
directions that may vary with dimension.

\begin{assumption}[One-component distributional conditions]
\label{assumption:one-inference}
In addition to Assumptions~\ref{assumption:oracle-moments}
and~\ref{assumption:one-scale}, the common error distribution satisfies
\[
 0<c_\tau\le \mathbb{E}(\varepsilon^2)\le C_\tau<\infty,
 \qquad \mathbb{E}|\varepsilon|^4\le C_4.
\]
For some \(\delta_{\mathrm{mom}}>0\), the density ratio satisfies the uniform
source-domain moment bound
\[
 \sup_n \mathbb{E}_{\source}\{w_0(\boldsymbol{X})^{2+\delta_{\mathrm{mom}}}\}<\infty.
\]
\end{assumption}

A first-order expansion of the scalar ratio coefficient map yields the
mean-zero oracle influence function \(\boldsymbol{\phi}_{o,n}(\boldsymbol{X},Y)\in\mathbb R^p\) in
Supplementary Eq.~(56).
Supplementary Lemma~6 establishes the corresponding linear
representation of the projected coefficient error.
For a deterministic nonzero direction sequence \(\boldsymbol{v}_n\in\mathbb R^p\), define
\[
 s_{o,n}(\boldsymbol{v}_n)
 =\left[\frac1n\operatorname{Var}_{\source}
   \{\boldsymbol{v}_n^\top\boldsymbol{\phi}_{o,n}(\boldsymbol{X},Y)\}\right]^{1/2}.
\]

\begin{theorem}[Oracle projected asymptotic normality]
\label{thm:oracle-one-component-clt}
Under the model in \eqref{eq:theory-model} and
Assumptions~\ref{assumption:oracle-moments},
\ref{assumption:one-scale}, and~\ref{assumption:one-inference},
for every deterministic sequence of nonzero directions \(\boldsymbol{v}_n\),
\[
 \frac{
 \boldsymbol{v}_n^\top\bigl(\widetilde{\boldsymbol{\beta}}_1^{\mathrm{IWPLS}}
          -\boldsymbol{\beta}_{\target,1}^{\mathrm{PLS}}\bigr)
 }{s_{o,n}(\boldsymbol{v}_n)}
 \ \xrightarrow{d}\ N(0,1).
\]
\end{theorem}

The population standard error $s_{o,n}(\boldsymbol{v}_n)$
accounts for the variability induced by the true
importance weights. Supplementary Lemma~5  establishes its positivity
and directional variance bounds, while Supplementary Lemmas~6 and~7
control the standardized linearization remainder and
verify the Lyapunov condition, respectively.
These results justify the normal limit along
deterministic directions even when $p$ diverges.
The joint extension to a fixed number of
deterministic contrasts is established in
Supplementary Theorem~5 under its
nonsingularity condition.

\subsubsection{Estimated-Weight Asymptotic Distribution}

Estimating the importance weights introduces additional
source- and target-domain influence-function contributions.
Let \(\mathcal D_n\) denote the derivative of the
one-component coefficient map at the target moments,
defined in Supplementary Eq.~(52).
For a deterministic nonzero direction sequence
\(\boldsymbol{v}_n\), let
\(\gamma_a=\gamma_{a,n,v}\), \(a\in\{\source,\target\}\),
denote the deterministic, mean-zero, covariate-dependent
influence functions specified in
Assumption~\ref{assumption:estimated-inference}.
Define the combined influence functions and their
population standard error by
\[
\begin{aligned}
\psi_{\source}(\boldsymbol{X},Y)
&=\boldsymbol{v}_n^\top
\boldsymbol{\phi}_{o,n}(\boldsymbol{X},Y)
+\gamma_{\source}(\boldsymbol{X}),\\
\psi_{\target}(\boldsymbol{X})
&=\gamma_{\target}(\boldsymbol{X}),\\
s_{\mathrm{IW},n}^2(\boldsymbol{v}_n)
&=\frac{
\operatorname{Var}_{\source}(\psi_{\source})
+\operatorname{Var}_{\target}(\psi_{\target})
}{n}.
\end{aligned}
\]

The following assumption requires the weight-induced
moment perturbation to admit a first-order expansion
with a negligible remainder on this standard-error
scale. It also imposes a standardized moment condition
on the additional influence functions.

\begin{assumption}[Estimated-weight distributional conditions]
\label{assumption:estimated-inference}
Let \(a_n=n^{-1/2}+r_w\).
Suppose that
\(\sqrt{n}a_n^2\to0\) and
\(0<s_{\mathrm{IW},n}^2(\boldsymbol{v}_n)<\infty\).
For the specified deterministic direction sequence
\(\boldsymbol{v}_n\), assume
\begin{equation}
\begin{aligned}
&\boldsymbol{v}_n^\top\mathcal D_n
\left(
\widehat{\boldsymbol{\Sigma}}
-\widetilde{\boldsymbol{\Sigma}},
\widehat{\boldsymbol{\sigma}}
-\widetilde{\boldsymbol{\sigma}}
\right)\\
&\quad=
\frac1n\sum_{i=1}^{n}
\gamma_{\source,n,v}(\boldsymbol{X}_i^{\source})
+\frac1n\sum_{j=1}^{n}
\gamma_{\target,n,v}(\boldsymbol{X}_j^{\target})
+o_{\mathbb P}
\{s_{\mathrm{IW},n}(\boldsymbol{v}_n)\}.
\end{aligned}
\label{eq:estimated-weight-moment-linearization}
\end{equation}
Moreover, there exist
\(0<\eta_{\mathrm{mom}}\le\min\{\delta_{\mathrm{mom}}/2,1\}\) and
\(C_\gamma<\infty\), independent of \(n\),
such that, for \(a\in\{\source,\target\}\),
\begin{equation}
\mathbb{E}_a|\gamma_{a,n,v}|^{2+\eta_{\mathrm{mom}}}
\le C_\gamma
\{\operatorname{Var}_a(\gamma_{a,n,v})\}^{1+\eta_{\mathrm{mom}}/2}.
\label{eq:estimated-weight-influence-moments}
\end{equation}
\end{assumption}

The rate condition is equivalent to \(r_w=o(n^{-1/4})\) and
controls the nonlinear coefficient remainder.
Weight estimation can still contribute to the leading influence functions
under this rate.
The first-order expansion and standardized moment bound are separate
requirements. The expansion must include normalization and centering,
with any remaining regularization or representation bias negligible
relative to \(s_{\mathrm{IW},n}(\boldsymbol{v}_n)\).

\begin{theorem}[Estimated-weight projected asymptotic normality]
\label{thm:estimated-one-component-clt}
Under the model in \eqref{eq:theory-model} and
Assumptions~\ref{assumption:oracle-moments},
\ref{assumption:estimated-weight}, \ref{assumption:one-scale},
\ref{assumption:one-inference}, and~\ref{assumption:estimated-inference},
\[
 \frac{
 \boldsymbol{v}_n^\top(\widehat{\boldsymbol{\beta}}_1^{\mathrm{IWPLS}}
          -\boldsymbol{\beta}_{\target,1}^{\mathrm{PLS}})
 }{s_{\mathrm{IW},n}(\boldsymbol{v}_n)}
 \ \xrightarrow{d}\ N(0,1).
\]
\end{theorem}

The source variance decomposition in
Supplementary Eq.~(87) accounts for
the covariance between the oracle and weight-estimation
contributions, which may have either sign. Thus, weight
estimation need not increase the source variance, while
the independent target sample contributes an additional
variance component. Both normal limits use population
standard errors, and comparing the estimators requires
control of their projected difference on the corresponding
standard-error scale.

Theorem~3 requires estimator-specific verification of
Assumption~5 beyond weight consistency.
Supplementary Section~I.3
provides sufficient conditions for correctly specified
fixed-basis uLSIF, establishing a two-sample influence-function
expansion and a negligible standardized remainder even
when source observations are reused. 

\subsection{Multi-Component IWPLS: Consistency}
\label{sec:theory-multi}

For fixed \(K\) in the component range above, we impose a geometry
condition that separates the Krylov column scales from stability of the
reduced regression problem,
following \citet{CookForzani2019}. Recall
\[
 \mathbf{G}_{\target,K}=(\boldsymbol{\sigma}_{\target},\boldsymbol{\Sigma}_{\target}\boldsymbol{\sigma}_{\target},\ldots,
 \boldsymbol{\Sigma}_{\target}^{K-1}\boldsymbol{\sigma}_{\target}),\qquad
 \boldsymbol{\Theta}_{\target,K}=\mathbf{G}_{\target,K}^\top\boldsymbol{\Sigma}_{\target}\mathbf{G}_{\target,K}.
\]
Define
\[
 m_\ell=\boldsymbol{\sigma}_{\target}^\top\boldsymbol{\Sigma}_{\target}^\ell\boldsymbol{\sigma}_{\target},
 \qquad 0\le\ell\le2K-1,
\]
and, for positive \(m_1,m_3,\ldots,m_{2K-1}\), set
\[
 \mathbf{D}_{\target,K}=\operatorname{diag}(m_1,m_3,\ldots,m_{2K-1}),\qquad
 \mathbf{C}_{\target,K}=\mathbf{D}_{\target,K}^{-1/2}\boldsymbol{\Theta}_{\target,K}\mathbf{D}_{\target,K}^{-1/2}.
\]
Thus \(m_0=q_{\target}\) and \(m_1=m_{\target}\).
The matrices \(\mathbf{G}_{\target,K}\in\mathbb R^{p\times K}\) and
\(\boldsymbol{\Theta}_{\target,K}\in\mathbb R^{K\times K}\) define the reduced system.
The matrix \(\mathbf{C}_{\target,K}\) is the Gram matrix of the covariance-norm
normalized Krylov columns, with unit diagonal.

\begin{assumption}[Fixed-component Krylov geometry]
\label{assumption:fixed-d}

Fix an integer $K\ge1$.
There exist constants
$0<c_m\le C_m<\infty$ and $c_C>0$,
independent of $n$ and $p$, such that
\[
c_m p^{\ell+1}
\le m_\ell
\le C_m p^{\ell+1},
\qquad
0\le\ell\le2K-1,
\]
and
\[
\lambda_{\min}
(\mathbf{C}_{\target,K})
\ge c_C.
\]
\end{assumption}

The moment bounds control the
Krylov column scales, while the
eigenvalue condition ensures
stability of the normalized
reduced Gram matrix.
Since
$\mathbf{C}_{\target,K}\succ0$,
the first $K$ target Krylov vectors
are linearly independent.
Consequently, $K\le d_{\target}$.
For $K=1$, the condition reduces
to Assumption~\ref{assumption:one-scale}.

Combining the oracle and estimated-weight moment bounds in
Supplementary Section~D with this geometry condition yields the
following consistency rates under their respective assumptions.
The normalized sample Gram perturbations in Supplementary Lemma~11
ensure nonsingularity of the sample reduced systems in
Section~\ref{sec:iwpls-methodology}.

\begin{theorem}[Fixed-component consistency]
\label{thm:fixed-d-consistency}
Fix \(1\le K<\min\{d_{\source},d_{\target}\}\).
Under Assumptions~\ref{assumption:oracle-moments} and~\ref{assumption:fixed-d},
\[
 \|\widetilde{\boldsymbol{\beta}}_K^{\mathrm{IWPLS}}-\boldsymbol{\beta}_{\target,K}^{\mathrm{PLS}}\|_{\boldsymbol{\Sigma}_{\target}}
 =O_{\mathbb P}(n^{-1/2}).
\]
If Assumption~\ref{assumption:estimated-weight} also holds, then
\[
 \|\widehat{\boldsymbol{\beta}}_K^{\mathrm{IWPLS}}-\boldsymbol{\beta}_{\target,K}^{\mathrm{PLS}}\|_{\boldsymbol{\Sigma}_{\target}}
 =O_{\mathbb P}(n^{-1/2}+r_w).
\]
Under their respective assumptions, the oracle and estimated-weight
reduced Gram matrices are nonsingular with probability tending to one.
These conclusions also hold for \(K=1\), recovering
Theorem~\ref{thm:one-consistency}.
\end{theorem}

The estimated-weight bound retains the oracle rate when
\(r_w=O(n^{-1/2})\) and reduces to \(O_{\mathbb P}(r_w)\) when
\(n^{-1/2}=o(r_w)\).
The proof also gives
\(\|\widehat{\boldsymbol{\beta}}_K^{\mathrm{IWPLS}}-
\widetilde{\boldsymbol{\beta}}_K^{\mathrm{IWPLS}}\|_{\boldsymbol{\Sigma}_{\target}}
=O_{\mathbb P}(r_w)\).
Thus \(r_w=o(n^{-1/2})\) makes this difference negligible relative to
the oracle covariance-norm rate.
These are estimation rates for \(\boldsymbol{\beta}_{\target,K}^{\mathrm{PLS}}\).
Section~\ref{sec:theory-prediction} addresses the approximation error
relative to \(\boldsymbol{\beta}_0\).

\subsection{Target Prediction and Approximation Error}
\label{sec:theory-prediction}

Let \(\mathcal F_n\) denote the \(\boldsymbol{\sigma}\)-field generated by all data
and randomness used in density-ratio estimation and IWPLS fitting.
Let \(\boldsymbol{X}_{\mathrm{new}}^{\target}\sim P_{\target}^X\) be independent of
\(\mathcal F_n\).
To isolate the prediction error attributable to slope estimation
relative to the target-domain population PLS coefficient, define
\begin{equation}
 D_{\mathrm{new},K}
 =(\boldsymbol{X}_{\mathrm{new}}^{\target}-\boldsymbol{\mu}_{\target})^\top
 \bigl(\widehat{\boldsymbol{\beta}}_K^{\mathrm{IWPLS}}
       -\boldsymbol{\beta}_{\target,K}^{\mathrm{PLS}}\bigr).
 \label{eq:theory-slope-prediction-error}
\end{equation}
The oracle counterpart is obtained by replacing
\(\widehat{\boldsymbol{\beta}}_K^{\mathrm{IWPLS}}\) with
\(\widetilde{\boldsymbol{\beta}}_K^{\mathrm{IWPLS}}\).
Independence of \(\boldsymbol{X}_{\mathrm{new}}^{\target}\) from \(\mathcal F_n\) gives
\begin{equation}
 \mathbb{E}\!\left(D_{\mathrm{new},K}^2\mid\mathcal F_n\right)
 =
 \left\|\widehat{\boldsymbol{\beta}}_K^{\mathrm{IWPLS}}
       -\boldsymbol{\beta}_{\target,K}^{\mathrm{PLS}}\right\|_{\boldsymbol{\Sigma}_{\target}}^2.
 \label{eq:theory-prediction-norm-identity}
\end{equation}
The target covariance norm therefore equals the conditional root mean
squared prediction error due to slope estimation.
The complete prediction rule additionally depends on the estimated
weighted predictor and response means.

The target-domain population PLS rule is
\[
 f_{\target,K}^{\mathrm{PLS}}(\boldsymbol{x})=\nu_{\target}
 +(\boldsymbol{x}-\boldsymbol{\mu}_{\target})^\top\boldsymbol{\beta}_{\target,K}^{\mathrm{PLS}}.
\]
The coefficient analysis of Theorem~\ref{thm:fixed-d-consistency} and
the weighted-mean bounds in Supplementary Section~D yield the following
rates for the complete prediction rules.

\begin{corollary}[Fixed-component target prediction]
\label{cor:fixed-component-prediction}
Fix \(1\le K<\min\{d_{\source},d_{\target}\}\).
Under Assumptions~\ref{assumption:oracle-moments}
and~\ref{assumption:fixed-d}, let \(\boldsymbol{X}_{\mathrm{new}}^{\target}\) be a
new target-domain covariate independent of the fitting and
weight-estimation samples. Then
\[
 \widetilde f_K^{\mathrm{IWPLS}}(\boldsymbol{X}_{\mathrm{new}}^{\target})
 -f_{\target,K}^{\mathrm{PLS}}(\boldsymbol{X}_{\mathrm{new}}^{\target})
 =O_{\mathbb P}(n^{-1/2}).
\]
If Assumption~\ref{assumption:estimated-weight} also holds, then
\[
 \widehat f_K^{\mathrm{IWPLS}}(\boldsymbol{X}_{\mathrm{new}}^{\target})
 -f_{\target,K}^{\mathrm{PLS}}(\boldsymbol{X}_{\mathrm{new}}^{\target})
 =O_{\mathbb P}(n^{-1/2}+r_w).
\]
\end{corollary}

The corollary controls prediction error at an independent target
covariate relative to the target population PLS rule, excluding
irreducible response noise.
Integrating over \(P_{\target}^X\) with the fitted rules held fixed gives
the conditional mean squared errors in
Supplementary Eq.~(111).
These \(\mathcal F_n\)-measurable risks are \(O_{\mathbb P}(n^{-1})\)
for oracle weights and \(O_{\mathbb P}\{(n^{-1/2}+r_w)^2\}\) for estimated
weights under the respective conditions.

The target population PLS rule can differ from the true conditional mean
\(m(\boldsymbol{x})=\alpha_0+\boldsymbol{x}^\top\boldsymbol{\beta}_0\) under \eqref{eq:theory-model}.
The finite-component approximation error satisfies
\[
 f_{\target,K}^{\mathrm{PLS}}(\boldsymbol{x})-m(\boldsymbol{x})
 =(\boldsymbol{x}-\boldsymbol{\mu}_{\target})^\top
   (\boldsymbol{\beta}_{\target,K}^{\mathrm{PLS}}-\boldsymbol{\beta}_0).
\]
The root mean squared approximation error under \(P_{\target}^X\) is
\(\|\boldsymbol{\beta}_{\target,K}^{\mathrm{PLS}}-\boldsymbol{\beta}_0\|_{\boldsymbol{\Sigma}_{\target}}\).

Under the respective assumptions of
Corollary~\ref{cor:fixed-component-prediction}, the conditional root mean
squared estimation error relative to \(f_{\target,K}^{\mathrm{PLS}}\)
vanishes in probability by Supplementary Eq.~(111).
The triangle and reverse triangle inequalities therefore show that the
conditional root mean squared error relative to \(m\) vanishes in
probability if and only if
\(\|\boldsymbol{\beta}_{\target,K}^{\mathrm{PLS}}-\boldsymbol{\beta}_0\|_{\boldsymbol{\Sigma}_{\target}}\to0\)
(Supplementary Corollary~2).
For fixed \(K<d_{\target}\), this approximation error may persist even
when the target PLS functional is consistently estimated.

The target-domain approximation error is distinct from PLS-specific
PTD, which measures the difference between source- and target-domain
population PLS coefficients (Section~\ref{sec:target-pls-mismatch}).
For \(K=1<\min\{d_{\source},d_{\target}\}\), both domains'
population one-component PLS coefficients differ from \(\boldsymbol{\beta}_0\).
This condition alone does not imply a nonzero PLS-specific PTD,
which depends on the domain-specific moments

\section{Simulation Study}
We conduct a simulation study to evaluate working linear predictors
under joint covariate shifts and a common nonlinear conditional
response law across the source and target populations. For any
positive shift (\(\delta>0\)), our data-generating process
simultaneously alters the covariate mean vectors, marginal variances,
and off-diagonal covariances. The covariance shift alters the
predictor covariance matrix and the predictor--response cross-covariance.
The mean shift further affects the cross-covariance through the nonlinear
response term, so both population moments governing PLS component
extraction differ across domains. We assess the effect of density-ratio
weighting on PLS and compare supervised component extraction with ridge
regularization using the same oracle or estimated weights. Across all
tested positive shift levels, both oracle- and estimated-weight IWPLS
achieve higher mean target-domain \(Q^2\) and lower mean RMSEP than
unweighted PLS.

\subsection{Simulation Design}

The covariance design adapts the isotropic predictor-variation model of
\citet{CookForzani2019}, with $p=1000$ predictors and a common
covariance-invariant subspace of dimension $d=128$. Each replication
generates three independent samples of size 800: a labeled source
sample, an unlabeled target-adaptation sample, and a target-test
sample whose responses are used solely for out-of-sample evaluation.
Thus, the predictor dimension exceeds all three sample sizes.

Let $\mathbf{H}_p\in\mathbb{R}^{p\times d}$ have orthonormal columns,
with entries
\[
 (\mathbf{H}_p)_{j0}=p^{-1/2},\qquad
 (\mathbf{H}_p)_{jk}=\sqrt{\frac{2}{p}}
 \cos\left\{\frac{\pi(j+1/2)k}{p}\right\},
\]
for $j=0,\ldots,p-1$ and $k=1,\ldots,d-1$.
For $\rho\in(0,1)$ and $k=0,\ldots,d-1$, define
\[
 \begin{gathered}
 g_k=16^{1-k/(d-1)},\qquad
 t_k=\frac{5p g_k}{\sum_{l=0}^{d-1}g_l},\qquad
 b_\rho=\frac{5(1-\rho)}{\rho},\\
 \mathbf{D}_p=\operatorname{diag}(\sqrt{t_0},\ldots,\sqrt{t_{d-1}}),
 \qquad \boldsymbol{\Theta}_p=\mathbf{H}_p\mathbf{D}_p,\\
 \boldsymbol{\Lambda}_\rho=\mathbf{D}_p^2+b_\rho\mathbf{I}_d,
 \qquad
 \boldsymbol{\Sigma}_\source
 =\boldsymbol{\Theta}_p\boldsymbol{\Theta}_p^\top+b_\rho\mathbf{I}_p.
 \end{gathered}
\]
The nonzero eigenvalues $t_0,\ldots,t_{d-1}$ of the factor component
$\boldsymbol{\Theta}_p\boldsymbol{\Theta}_p^\top$ decrease
geometrically with a largest-to-smallest ratio of 16 and sum to $5p$.
Consequently, $\rho$ is the source factor variance share,
\[
 \rho=\frac{\operatorname{tr}
 (\boldsymbol{\Theta}_p\boldsymbol{\Theta}_p^\top)}
 {\operatorname{tr}(\boldsymbol{\Sigma}_\source)},
\]
which represents the proportion of total source predictor variance
attributable to the factor component. For $\rho=0.7,0.8,0.9$, and
$0.95$, the resulting mean correlations between adjacent source
predictors are approximately $0.691,0.790,0.889$, and $0.939$,
respectively.

Fix $\boldsymbol{v}=(v_0,\ldots,v_{d-1})^\top\in\mathbb{R}^d$ with
entries $v_k=s_k\sqrt{8/d}$, where the signs $s_k\in\{-1,1\}$ are
generated using random seed 20261001. Define the orthonormal
factor-space directions
\[
 \boldsymbol{u}
 =\frac{(v_1,-v_0,0,\ldots,0)^\top}{\sqrt{v_0^2+v_1^2}},
 \qquad \boldsymbol{e}=\frac{\boldsymbol{v}}{\sqrt{8}},
\]
and set
\[
 \boldsymbol{\beta}_\rho
 =\mathbf{H}_p\boldsymbol{\Lambda}_\rho^{-1/2}\boldsymbol{v},
 \qquad
 \boldsymbol{q}_\rho
 =\mathbf{H}_p\boldsymbol{\Lambda}_\rho^{1/2}\boldsymbol{u},
 \qquad
 \boldsymbol{a}_\rho
 =\mathbf{H}_p\boldsymbol{\Lambda}_\rho^{-1/2}\boldsymbol{u}.
\]
By construction, these vectors satisfy
$\boldsymbol{\beta}_\rho^\top\boldsymbol{\Sigma}_\source
\boldsymbol{\beta}_\rho=8$,
$\boldsymbol{\beta}_\rho^\top\boldsymbol{q}_\rho=0$, and
$\boldsymbol{a}_\rho^\top\boldsymbol{q}_\rho=1$.
Each standardized source eigen-coordinate in the factor subspace
contributes $8/d$ to the linear signal variance. At least 99\% of the
entries of $\boldsymbol{\beta}_\rho$ are nonzero in the original predictor
coordinates for the fixed sign realization. For fixed $d$ and $\rho$,
the $d$ factor-subspace eigenvalues of $\boldsymbol{\Sigma}_\source$
are of order $p$, the remaining $p-d$ background eigenvalues equal
$b_\rho$, and $\|\boldsymbol{\beta}_\rho\|_2^2\asymp p^{-1}$. This
construction distributes response information across many predictors
while keeping the source linear signal variance fixed.

The shift parameter $\delta\ge0$ controls the joint shifts in the
covariate mean vectors and covariance matrices. Let
\[
 \begin{gathered}
 \mathbf{M}=\boldsymbol{e}\boldsymbol{e}^\top
           -\boldsymbol{u}\boldsymbol{u}^\top,
 \qquad \eta=\frac{\log(1.25)}{1.5},\\
 \mathbf{C}_\delta=\exp(\eta\delta\mathbf{M})
 =\mathbf{I}_d+(e^{-\eta\delta}-1)\boldsymbol{u}\boldsymbol{u}^\top
             +(e^{\eta\delta}-1)\boldsymbol{e}\boldsymbol{e}^\top,\\
 \boldsymbol{\Sigma}_\target(\delta)
 =\boldsymbol{\Theta}_p\mathbf{C}_\delta
  \boldsymbol{\Theta}_p^\top+b_\rho\mathbf{I}_p.
 \end{gathered}
\]
We generate the source and target covariates as
\[
 \boldsymbol{X}^\source\sim N_p\left(
 -\frac{\delta}{2}\boldsymbol{q}_\rho,\boldsymbol{\Sigma}_\source\right),
 \qquad
 \boldsymbol{X}^\target\sim N_p\left(
 +\frac{\delta}{2}\boldsymbol{q}_\rho,
 \boldsymbol{\Sigma}_\target(\delta)\right).
\]
At $\delta=0$, the two covariate distributions coincide. At
$\delta=1.5$, the eigenvalues of $\mathbf{C}_\delta$ associated with
$\boldsymbol{u}$ and $\boldsymbol{e}$ are $0.8$ and $1.25$,
respectively. For any $\delta>0$, the covariance difference
$\boldsymbol{\Sigma}_\target(\delta)-\boldsymbol{\Sigma}_\source
=\boldsymbol{\Theta}_p(\mathbf{C}_\delta-\mathbf{I}_d)
\boldsymbol{\Theta}_p^\top$ has rank two and alters both the marginal
variances and the off-diagonal covariances in the original predictor
coordinates. Both domains retain the same loading matrix
$\boldsymbol{\Theta}_p$, isotropic background $b_\rho\mathbf{I}_p$,
and factor subspace $\operatorname{span}(\mathbf{H}_p)$. At
$\rho=0.85$ and $\delta=1.5$, the mean correlations between adjacent
predictors are approximately $0.84$ in both domains. The target
covariance matrix is positive definite for every $\delta\ge0$.
Across the tested range of $\delta$, the oracle density ratio has a
finite second moment under the source distribution.

Both domains follow the same conditional response model
\[
 Y=\boldsymbol{X}^\top\boldsymbol{\beta}_\rho
 +\gamma\left\{(\boldsymbol{a}_\rho^\top\boldsymbol{X})^2-1\right\}
 +\varepsilon,\qquad
 \gamma=0.4,\qquad
 \varepsilon\sim N(0,0.5^2),\quad\varepsilon\perp\boldsymbol{X}.
\]
The quadratic term makes the working linear regression model misspecified.
For a Gaussian covariate vector with mean
$\boldsymbol{\mu}_\ell$ and covariance $\boldsymbol{\Sigma}_\ell$ in
domain $\ell\in\{\source,\target\}$, the slope of the population best
affine predictor and the predictor--response cross-covariance vector
satisfy
\[
 \boldsymbol{b}_\ell^*
 =\boldsymbol{\beta}_\rho
  +2\gamma(\boldsymbol{a}_\rho^\top\boldsymbol{\mu}_\ell)
   \boldsymbol{a}_\rho,
 \qquad
 \boldsymbol{\sigma}_\ell
 =\operatorname{Cov}_\ell(\boldsymbol{X},Y)
 =\boldsymbol{\Sigma}_\ell\boldsymbol{b}_\ell^*.
\]
Because $\boldsymbol{\mu}_\source=-\frac{\delta}{2}\boldsymbol{q}_\rho$,
$\boldsymbol{\mu}_\target=+\frac{\delta}{2}\boldsymbol{q}_\rho$, and
$\boldsymbol{a}_\rho^\top\boldsymbol{q}_\rho=1$, we have
\[
 \boldsymbol{b}_\source^*
 =\boldsymbol{\beta}_\rho-\gamma\delta\boldsymbol{a}_\rho,
 \qquad
 \boldsymbol{b}_\target^*
 =\boldsymbol{\beta}_\rho+\gamma\delta\boldsymbol{a}_\rho.
\]
Under this construction, every $\delta>0$ yields different best affine
predictors and predictor--response cross-covariance vectors across domains.
The subspace $\operatorname{span}(\mathbf{H}_p)$ contains
$\boldsymbol{\beta}_\rho$ and $\boldsymbol{a}_\rho$ and is invariant
under both domain covariance matrices. It therefore contains the
population PLS Krylov subspaces in both domains.

We compare unweighted PLS with IWPLS and ridge-regularized
importance-weighted least squares (Ridge-IWLS), evaluating both
weighted methods using oracle and estimated weights. For estimated
weights, we apply principal component analysis (PCA) followed by
unconstrained least-squares importance fitting (uLSIF) to estimate
density ratios, a procedure we refer to as PCA-uLSIF and detail in
Supplementary Section~A.1.1. The resulting five
methods are PLS, Oracle-IWPLS, PCA-uLSIF-IWPLS, Oracle-Ridge-IWLS, and
PCA-uLSIF-Ridge-IWLS.

The oracle methods use the exact Gaussian density ratio
$w_0(\boldsymbol{x})=p_\target(\boldsymbol{x})/p_\source(\boldsymbol{x})$,
which accounts for both the mean and covariance shifts. Because the
common Gaussian density on the orthogonal complement of the factor
subspace cancels in $w_0(\boldsymbol{x})$, the oracle ratio can be
evaluated directly from the factor-subspace projection
$\mathbf{H}_p^\top\boldsymbol{x}$. In contrast, PCA-uLSIF estimates a
density ratio in a data-driven PCA score space using only the source
and target-adaptation covariates, without access to the population
loadings or response directions.

All methods share the same shuffled five-fold partition of the source
sample for prediction-model tuning. Unweighted PLS uses standard
source cross-validation, whereas the weighted methods use the foldwise
self-normalized IWCV criterion in Supplementary Eq.~(39), averaged
equally across the five folds. The candidate PLS component counts are
$K\in\{1,\ldots,40\}$, where each candidate must be feasible in every
source-training fold and in the full-source fit. Ridge-IWLS is fitted
with an unpenalized intercept and training weights normalized to sum
to one, using candidate penalties
$\{10^{-4},10^{-3},10^{-2},10^{-1},1,10,100,1000\}$. Cross-validation
ties favor the smaller component count or penalty. Unweighted
PLS centers the predictors and response using unweighted
source-training means, whereas the weighted methods use
importance-weighted source-training means, with the fitted intercepts
retained for target-domain prediction.

Within each outer source-training fold, an inner five-fold
density-ratio cross-validation partitions both the outer
source-training covariates and the target-adaptation covariates to
jointly select the PCA dimension $L\in\{2,5,10,20\}$, the kernel
bandwidth multiplier $h\in\{0.5,1,2\}$, and the uLSIF penalty
$\lambda\in\{0.01,0.1,1\}$. In each inner split, we fit PCA to the
pooled inner-training source and target covariates and standardize
the resulting scores using their pooled training means and standard
deviations. We set the Gaussian kernel bandwidth to $h$ times the median
positive pairwise Euclidean distance among the pooled standardized
inner-training scores. We sample 20 kernel centers from the
standardized inner-training target scores. On the held-out
inner-validation source and target covariates, the unnormalized
nonnegative ratio estimates $r$ are evaluated using the empirical
uLSIF criterion
\[
 \frac12\operatorname{mean}_{\source,\mathrm{val}}(r^2)
 -\operatorname{mean}_{\target,\mathrm{val}}(r),
\]
averaged equally across the five inner folds.
Among candidates whose mean validation scores are within $10^{-12}$
of the minimum, we select the smallest $L$, then $h$, and then $\lambda$.

After selecting $(L,h,\lambda)$, we refit the density-ratio estimator
on the full outer source-training fold and all target-adaptation
covariates. The refitted estimator is also evaluated on the held-out
outer source-validation covariates, and both the training and
validation ratio estimates are normalized by the mean of the ratio
estimates over the outer source-training sample. Within each weight
type, IWPLS and Ridge-IWLS share the same weight functions. To obtain
the final weights, we repeat this density-ratio selection and
refitting procedure on all source and target-adaptation covariates.

Throughout, PCA is used solely for density-ratio estimation, whereas
all regression models are fitted on the $p$ original predictors. We
truncate negative uLSIF coefficients at zero and apply neither
upper-tail weight clipping nor filtering based on the effective
sample size (ESS). Within each outer split, the source-validation
observations are used only for held-out scoring, and the target-test
covariates and responses are reserved exclusively for final
prediction and evaluation, respectively.
Supplementary Section~A.2 details the complete nested
fitting procedure.

Experiment~A fixes $\delta=0.9$ and varies
$\rho\in\{0.7,0.8,0.9,0.95\}$, whereas Experiment~B fixes
$\rho=0.85$ and varies $\delta\in\{0,0.3,0.6,0.9,1.2,1.5\}$. Both
experiments compare all five methods over 100 Monte Carlo
replications per setting. Initialized from master seed 20260925, five
independent random number streams generate the three covariate
samples and the two response-noise vectors. Within each experiment,
we use common underlying random draws across settings. For each
setting and replication, all five methods share the same generated
samples and source cross-validation folds.

We evaluate out-of-sample target prediction using $Q^2_\target$ and
the root mean squared error of prediction (RMSEP). For the
$n_{\mathrm{test}}=800$ target-test observations
($i=1,\ldots,n_{\mathrm{test}}$), let $Y_i^\target$ denote the
observed response, $\widehat Y_i^\target$ the predicted response, and
$\overline Y_\target$ the target-test sample mean. Define
\begin{equation}
 \begin{gathered}
 Q^2_\target=1-
 \frac{\sum_i(Y_i^\target-\widehat Y_i^\target)^2}
      {\sum_i(Y_i^\target-\overline Y_\target)^2},\\
 \mathrm{RMSEP}_\target
 =\left\{\frac{1}{n_{\mathrm{test}}}
 \sum_i(Y_i^\target-\widehat Y_i^\target)^2\right\}^{1/2}.
 \end{gathered}
 \label{eq:simulation-q2}
\end{equation}
Larger $Q^2_\target$ and smaller RMSEP indicate better prediction
accuracy. Because the target response variance changes with $\delta$,
we interpret the two metrics jointly.
Figure~\ref{fig:simulation-target-prediction} displays both metrics
across all settings, and Table~\ref{tab:simulation-q2} reports the
Monte Carlo means and between-replication sample standard deviations
of $Q^2_\target$ over the 100 replications.

\subsection{IWPLS Improves Prediction across Factor Shares}

\begin{figure}[!t]
\centering
\includegraphics[width=\textwidth]{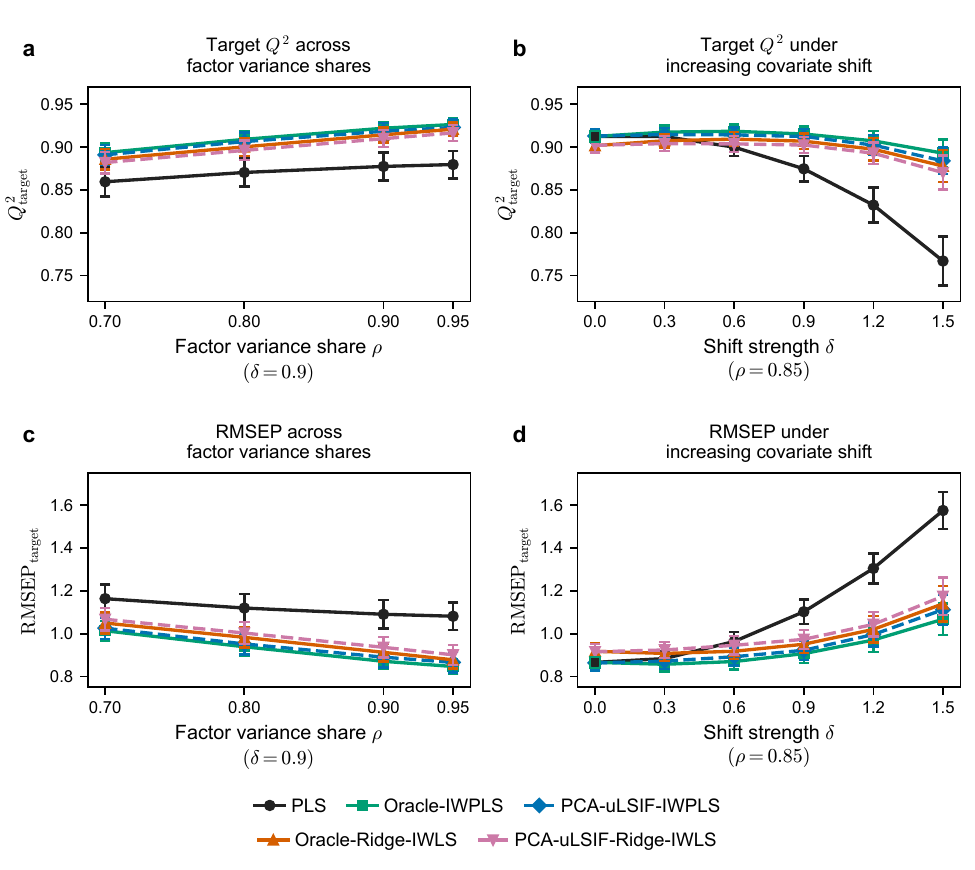}
\caption{Target prediction across source factor variance shares and
covariate shift strengths. Panels~(a) and~(c) vary $\rho$ at
$\delta=0.9$ (Experiment~A), whereas panels~(b) and~(d) vary $\delta$
at $\rho=0.85$ (Experiment~B). The upper and lower rows report
$Q^2_\target$ and $\mathrm{RMSEP}_\target$, respectively, with a
common vertical scale in each row. Solid lines denote PLS and the
oracle methods, and dashed lines denote the PCA-uLSIF methods. Points
represent Monte Carlo means over 100 replications per setting, and
error bars show one between-replication sample standard deviation
above and below each mean. At $\delta=0$, PLS and Oracle-IWPLS
coincide, with a green square outline marking their shared position.}
\label{fig:simulation-target-prediction}
\end{figure}

All five methods achieve higher mean $Q^2_\target$ and lower mean
$\mathrm{RMSEP}_\target$ as $\rho$ increases
(Figure~\ref{fig:simulation-target-prediction}, panels~(a) and~(c);
Table~\ref{tab:simulation-q2}, block~A). At every factor share,
Oracle-IWPLS attains the highest mean $Q^2_\target$ and the lowest mean
$\mathrm{RMSEP}_\target$. PCA-uLSIF-IWPLS follows closely, with mean
$Q^2_\target$ values $0.0026$--$0.0036$ lower than those of
Oracle-IWPLS across the four settings. Both IWPLS variants
outperform their Ridge-IWLS counterparts using the same weights in
both mean $Q^2_\target$ and mean $\mathrm{RMSEP}_\target$.

Across values of $\rho$, the source linear signal variance remains
fixed while the isotropic background variance $b_\rho$ decreases
as the factor share grows. The results support combining importance
weighting with supervised PLS component extraction in this
abundant-regression setting with correlated predictors. The
predictive gain over unweighted PLS persists when importance
weights are estimated from covariates using PCA-uLSIF.

\subsection{Weighting Limits PLS Deterioration under Shift}

At $\delta=0$, the true density ratio equals one, so PLS and
Oracle-IWPLS coincide in every replication, yielding identical
selected component counts, $Q^2_\target$, and
$\mathrm{RMSEP}_\target$. PCA-uLSIF-IWPLS attains a mean
$Q^2_\target$ of $0.9128$ compared with $0.9123$ for PLS and
Oracle-IWPLS (Table~\ref{tab:simulation-q2}, block~B). The small
difference is consistent with finite-sample variation in the
estimated weights.

As the shift strength $\delta$ increases, unweighted PLS loses target
prediction accuracy (Figure~\ref{fig:simulation-target-prediction},
panels~(b) and~(d); Table~\ref{tab:simulation-q2}, block~B).
Across all five positive shift levels, both IWPLS variants achieve
higher mean $Q^2_\target$ and lower mean $\mathrm{RMSEP}_\target$
than PLS. Oracle-IWPLS attains the highest mean $Q^2_\target$ and
lowest mean $\mathrm{RMSEP}_\target$ at every positive shift.
At $\delta=1.5$, PCA-uLSIF-IWPLS attains a mean $Q^2_\target$ of
$0.8838$ compared with $0.7671$ for PLS, with corresponding mean
$\mathrm{RMSEP}_\target$ values of $1.112$ and $1.575$. At this
shift level, Oracle-IWPLS reaches a mean $Q^2_\target$ of $0.8928$,
exceeding PCA-uLSIF-IWPLS by $0.0090$. At every shift level, both
IWPLS variants outperform their Ridge-IWLS counterparts using the
same weights in both mean $Q^2_\target$ and mean
$\mathrm{RMSEP}_\target$.

At $\delta=1.5$, the mean selected component counts are $8.15$,
$17.22$, and $12.24$ for PLS, Oracle-IWPLS, and PCA-uLSIF-IWPLS,
respectively. Across both experiments, none of the 3000 selected
component counts for these three methods reaches the upper limit of
$K=40$. For the final fits at $\delta=1.5$, the mean sample ESS
fractions are $0.1599$ for the oracle weights and $0.4554$ for the
estimated weights. Because ESS measures weight concentration, a larger
value alone does not imply more accurate density-ratio estimation.

Both IWPLS variants improve target prediction over unweighted PLS
across all five tested positive shift levels in Experiment~B.
The comparison between Oracle-IWPLS and PCA-uLSIF-IWPLS quantifies
the observed prediction gap between oracle and estimated weighting.
Under the same oracle or estimated weights, IWPLS achieves higher
mean $Q^2_\target$ and lower mean $\mathrm{RMSEP}_\target$ than
Ridge-IWLS across all tested settings.

\begin{table}[!htbp]
\setlength{\abovecaptionskip}{0pt}
\setlength{\belowcaptionskip}{6pt}
\centering
\caption{Target-domain $Q^2_\target$ across source factor variance shares and covariate shift strengths. Larger values indicate better target prediction.}
\label{tab:simulation-q2}
\begingroup
\small
\setlength{\tabcolsep}{4pt}
\renewcommand{\arraystretch}{1.1}
\begin{tabular}{@{}lccccc@{}}
\toprule
 & & \multicolumn{2}{c}{IWPLS} & \multicolumn{2}{c}{Ridge-IWLS} \\
\cmidrule(lr){3-4}\cmidrule(lr){5-6}
Setting & PLS & Oracle & PCA-uLSIF & Oracle & PCA-uLSIF \\
\midrule
\multicolumn{6}{@{}l}{\textit{A. Source factor variance shares ($\delta=0.9$)}} \\
\addlinespace[2pt]
$\rho=0.70$ & $0.8595\;(0.0174)$ & $\mathbf{0.8935}\;(0.0106)$ & $0.8909\;(0.0115)$ & $0.8858\;(0.0119)$ & $0.8819\;(0.0126)$ \\
$\rho=0.80$ & $0.8704\;(0.0166)$ & $\mathbf{0.9090}\;(0.0089)$ & $0.9063\;(0.0104)$ & $0.9002\;(0.0104)$ & $0.8961\;(0.0113)$ \\
$\rho=0.90$ & $0.8774\;(0.0162)$ & $\mathbf{0.9219}\;(0.0073)$ & $0.9183\;(0.0094)$ & $0.9142\;(0.0087)$ & $0.9097\;(0.0100)$ \\
$\rho=0.95$ & $0.8796\;(0.0161)$ & $\mathbf{0.9263}\;(0.0069)$ & $0.9229\;(0.0090)$ & $0.9208\;(0.0080)$ & $0.9165\;(0.0092)$ \\
\addlinespace[3pt]
\midrule
\multicolumn{6}{@{}l}{\textit{B. Covariate shift strengths ($\rho=0.85$)}} \\
\addlinespace[2pt]
$\delta=0.0$ & $0.9123\;(0.0077)$ & $0.9123\;(0.0077)$ & $\mathbf{0.9128}\;(0.0078)$ & $0.9016\;(0.0087)$ & $0.9020\;(0.0086)$ \\
$\delta=0.3$ & $0.9122\;(0.0082)$ & $\mathbf{0.9174}\;(0.0075)$ & $0.9146\;(0.0077)$ & $0.9074\;(0.0081)$ & $0.9040\;(0.0086)$ \\
$\delta=0.6$ & $0.8999\;(0.0105)$ & $\mathbf{0.9184}\;(0.0080)$ & $0.9142\;(0.0085)$ & $0.9092\;(0.0081)$ & $0.9036\;(0.0091)$ \\
$\delta=0.9$ & $0.8746\;(0.0146)$ & $\mathbf{0.9152}\;(0.0089)$ & $0.9123\;(0.0088)$ & $0.9068\;(0.0089)$ & $0.9023\;(0.0092)$ \\
$\delta=1.2$ & $0.8323\;(0.0205)$ & $\mathbf{0.9072}\;(0.0116)$ & $0.9023\;(0.0118)$ & $0.8975\;(0.0133)$ & $0.8928\;(0.0125)$ \\
$\delta=1.5$ & $0.7671\;(0.0287)$ & $\mathbf{0.8928}\;(0.0160)$ & $0.8838\;(0.0154)$ & $0.8778\;(0.0188)$ & $0.8702\;(0.0199)$ \\
\bottomrule
\end{tabular}
\endgroup
\par\smallskip
\begin{minipage}{\linewidth}
\footnotesize
\textit{Notes.} Entries are Monte Carlo means with between-replication
sample standard deviations in parentheses, based on 100 replications
per setting.
Boldface indicates the largest unrounded Monte Carlo mean within each setting.
Oracle methods use exact target-to-source density ratios, whereas
PCA-uLSIF methods use estimated density ratios.
\end{minipage}
\end{table}

\FloatBarrier

\section{Empirical Studies under Covariate Shift}
\label{sec:real-data-evaluation}
\subsection{Data and Methods}

We evaluate IWPLS for target-domain prediction under covariate shift using two spectroscopic datasets previously studied in the context of covariate shift: the International Diffuse Reflectance Conference (IDRC) Shootout dataset~\citep{LarsenEtAl2020} and the Melamine dataset~\citep{NikzadLangerodiEtAl2020}. Both experiments use labeled source observations and unlabeled target covariates. Table~\ref{tab:empirical-tasks} lists the source and target subsets, sample sizes, and numbers of predictors.

\begin{table}[htbp]
\setlength{\abovecaptionskip}{0pt}
\setlength{\belowcaptionskip}{6pt}
\centering
\small
\caption{Source and target subsets.}
\label{tab:empirical-tasks}
\renewcommand{\arraystretch}{1.2}
\setlength{\tabcolsep}{9pt}
\begin{tabular}{@{}lccrrr@{}}
\toprule
Dataset & Source & Target
& \(n_{\source}\) & \(n_{\target}\) & \(p\) \\
\midrule
IDRC
& \textnormal{calibrate\_2}
& \textnormal{test\_2}
& 155 & 460 & 650 \\

Melamine
& \textnormal{R861}
& \textnormal{R568}
& 3890 & 733 & 346 \\
\bottomrule
\end{tabular}
\end{table}

We compare the target-domain prediction performance of ordinary PLS, PCA-uLSIF-IWPLS, and PCA-uLSIF-Ridge-IWLS across the two real-data tasks. Both weighted methods estimate source-sample importance weights using PCA-uLSIF based on source and unlabeled target covariates. They use the same estimated weights within each source-training fold and in the final fit, allowing their prediction performance to be compared under a common weighting scheme.

For the high-dimensional IDRC task, we additionally include Logistic-IWPLS as a sensitivity analysis of the density-ratio estimation procedure. It estimates source weights using an \(L_2\)-regularized logistic domain classifier, providing an alternative to PCA-uLSIF based on domain classification. This additional comparison examines whether the prediction gains of IWPLS are maintained when a different density-ratio estimator is used.

All methods use standard normal variate (SNV) preprocessing, a common scatter-correction procedure in near-infrared spectroscopy that has also been employed with PLS regression in pharmaceutical tablet calibration studies~\citep{ChalusEtAl2005,MishraEtAl2021}. For IDRC, SNV is applied separately to each spectrum, whereas for Melamine, it is applied separately to each spectral block before concatenation. Density-ratio estimation and component selection follow the procedures in Supplementary Section~A.

Within each experiment, all methods share the same fixed five-fold partition of the source sample. Regression tuning uses unweighted source validation for ordinary PLS and the self-normalized criterion in Supplementary Eq.~(39) for the weighted methods. After tuning, we refit the selected regression models using all labeled source observations and re-estimate the weights for the weighted methods using all source and unlabeled target covariates. Target responses are withheld from weight estimation, tuning, and model fitting and used only for final prediction evaluation. We report the root mean squared error of prediction (RMSEP), mean absolute error (MAE), and target-domain \(Q^2\) defined in \eqref{eq:simulation-q2}. We also report the effective sample size (ESS) ratio to summarize weight concentration.

\subsection{IDRC Shootout Data}

For the IDRC dataset~\citep{Hopkins2003Shootout}, we predict the
laboratory-measured active ingredient content (\textnormal{assay}) using
\textnormal{calibrate\_2} as the source sample and \textnormal{test\_2} as the
target sample.

We report the effective sample size (ESS) ratio to quantify the concentration
of the estimated importance weights and the retained source-sample information.
PCA-uLSIF and Logistic-IWPLS yield ESS ratios of 0.837 and 0.742, respectively,
indicating less concentrated weights under PCA-uLSIF in this task.

The results in Table~\ref{tab:idrc-results} show that PCA-uLSIF-IWPLS selects a
more parsimonious model and achieves substantially better target-domain
prediction than ordinary PLS across all three evaluation metrics. Specifically,
PCA-uLSIF-IWPLS requires only \(K=3\) latent components compared with \(K=6\)
for ordinary PLS, while reducing RMSEP from \(5.587\) to \(4.988\) (a relative
reduction of approximately \(10.7\%\)) and increasing \(Q^2\) from \(0.8741\)
to \(0.8997\). Although both PCA-uLSIF-weighted methods improve upon unweighted
PLS across all three metrics, IWPLS further outperforms Ridge-IWLS under the
same estimated weights, attaining lower RMSEP and MAE and higher \(Q^2\) and
thus demonstrating a predictive advantage over importance-weighted ridge
regression in this task.

\begin{table}[htbp]
\setlength{\abovecaptionskip}{0pt}
\setlength{\belowcaptionskip}{6pt}
\centering
\small
\caption{Target-domain prediction results for the IDRC dataset.}
\label{tab:idrc-results}
\renewcommand{\arraystretch}{1.2}
\setlength{\tabcolsep}{8pt}
\begin{tabular}{@{}lrrrrr@{}}
\toprule
Method & \(K\) & Ridge \(\alpha\) & \(Q^2\) & RMSEP & MAE \\
\midrule
PLS
& 6 & --
& 0.8741 & 5.587 & 3.703 \\

PCA-uLSIF-IWPLS
& 3 & --
& 0.8997 & 4.988 & 3.203 \\

PCA-uLSIF-Ridge-IWLS
& -- & 0.01
& 0.8926 & 5.159 & 3.378 \\

Logistic-IWPLS
& 4 & --
& 0.9151 & 4.587 & 2.864 \\
\bottomrule
\end{tabular}
\end{table}

Using weights estimated by an \(L_2\)-regularized logistic domain classifier,
Logistic-IWPLS selects \(K=4\) components and attains an RMSEP of \(4.587\),
outperforming both ordinary PLS and PCA-uLSIF-IWPLS across all three target
prediction metrics. This confirms that the predictive gains of IWPLS in the
IDRC task are not specific to PCA-uLSIF and can also be achieved with a
classifier-based density-ratio estimator.
\FloatBarrier

\subsection{Melamine Data}

For the Melamine dataset~\citep{NikzadLangerodiEtAl2018,NikzadLangerodiEtAl2020},\footnote{Dataset
and documentation are available at
\href{https://github.com/RNL1/Melamine-Dataset}{the Melamine dataset
repository}.} we use \textnormal{R861} as the source sample and
\textnormal{R568} as the target sample. The response variable is the turbidity
point, the temperature at which the condensate first becomes turbid in an
offline test. Each record contains in-line Fourier-transform near-infrared
(FT-NIR) absorbance measurements across two spectral blocks of 225 and 121
channels, yielding 346 predictors in total.

We again report the ESS ratio to quantify the concentration of the estimated
importance weights and the retained source-sample information. For Melamine,
the ESS is \(236.36\), or \(6.08\%\) of the source sample, which is
substantially lower than the ESS ratio of \(0.837\) for IDRC and indicates
considerably greater weight concentration in the Melamine task.

Table~\ref{tab:melamine-results} shows that PCA-uLSIF-IWPLS achieves a
substantial improvement in target-domain prediction over ordinary PLS on the
Melamine dataset. Both methods select \(K=10\), the largest value in the
candidate set; without increasing the number of latent components, IWPLS
reduces RMSEP by \(56.2\%\) from \(8.002\) to \(3.504\) and increases \(Q^2\)
from \(0.7474\) to \(0.9516\). Although both PCA-uLSIF-weighted methods improve
upon ordinary PLS across all three prediction metrics, IWPLS outperforms
Ridge-IWLS (which selects the smallest candidate penalty, \(\alpha=10^{-4}\))
under the same estimated weights, highlighting the predictive advantage of the
PLS-based fitting procedure over ridge regression under a common
importance-weighting scheme.

\begin{table}[htbp]
\setlength{\abovecaptionskip}{0pt}
\setlength{\belowcaptionskip}{6pt}
\centering
\small
\caption{Target-domain prediction results for the Melamine dataset.}
\label{tab:melamine-results}
\renewcommand{\arraystretch}{1.2}
\setlength{\tabcolsep}{8pt}
\begin{tabular}{@{}lrrrrr@{}}
\toprule
Method & \(K\) & Ridge \(\alpha\) & \(Q^2\) & RMSEP & MAE \\
\midrule
PLS
& 10 & --
& 0.7474 & 8.002 & 7.670 \\

PCA-uLSIF-IWPLS
& 10 & --
& 0.9516 & 3.504 & 2.896 \\

PCA-uLSIF-Ridge-IWLS
& -- & \(10^{-4}\)
& 0.9242 & 4.383 & 3.379 \\
\bottomrule
\end{tabular}
\end{table}

\begin{figure}[!ht]
\centering
\includegraphics[width=\textwidth]{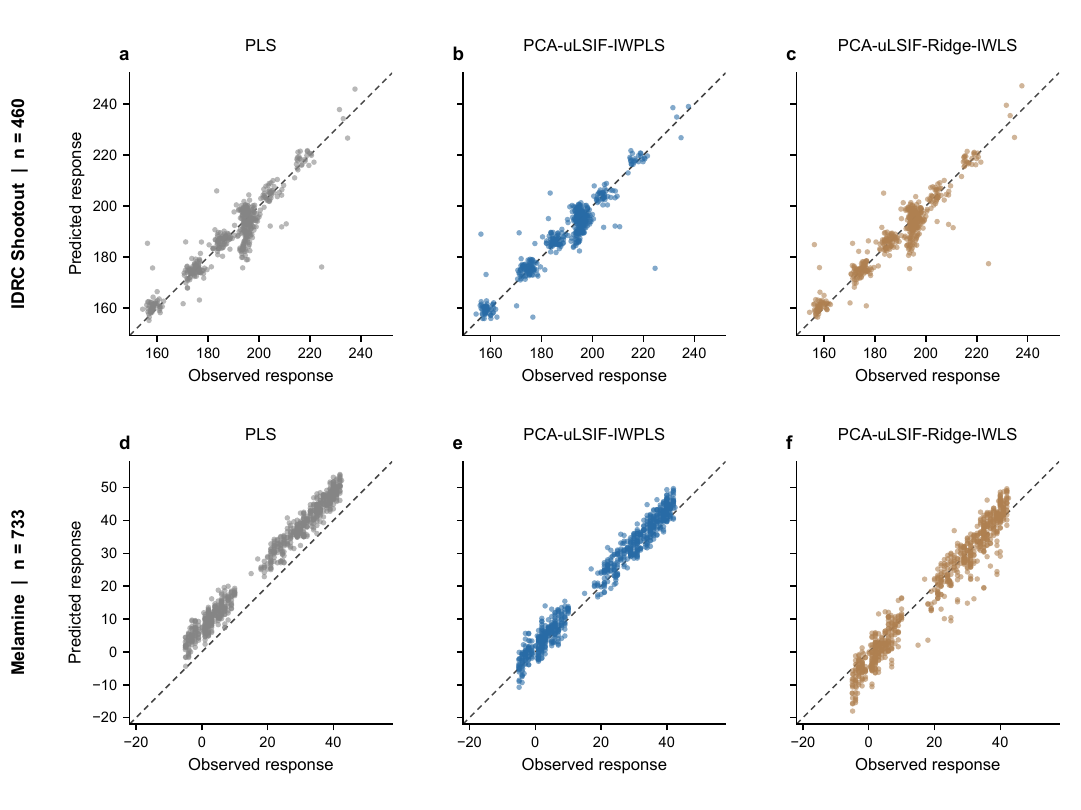}
\caption{Observed versus predicted responses for the target samples in IDRC (top
row, \(n=460\)) and Melamine (bottom row, \(n=733\)). From left to right, the
columns correspond to PLS, PCA-uLSIF-IWPLS, and PCA-uLSIF-Ridge-IWLS; the
dashed line indicates perfect prediction (\(\widehat{Y}=Y\)).}
\label{fig:empirical-predictions}
\end{figure}

Figure~\ref{fig:empirical-predictions} displays the target predictions,
showing the reduction in overprediction on Melamine with PCA-uLSIF-IWPLS.

Across both the IDRC and Melamine datasets, PCA-uLSIF-IWPLS consistently
outperforms ordinary PLS in target-domain prediction, achieving higher \(Q^2\)
as well as lower RMSEP and MAE.

\FloatBarrier

\section{Discussion}
This paper develops IWPLS to estimate the fixed-\(K\) target-population PLS
coefficient \(\boldsymbol{\beta}_{\target,K}^{\mathrm{PLS}}\) under covariate
shift using labeled source observations and unlabeled target covariates. By
reweighting the empirical means, predictor covariance, and predictor--response
cross-covariance without requiring target responses, IWPLS aligns both latent
component construction and regression fitting with the target-domain PLS
problem. Under a Gaussian linear model that accommodates strongly correlated
predictors and dense regression coefficients, IWPLS is consistent in the target
covariance norm as \(n,p\to\infty\) at the rate \(O_{\mathbb P}(n^{-1/2}+r_w)\),
retaining the oracle \(n^{-1/2}\) rate whenever the empirical weight-estimation
error satisfies \(r_w=O(n^{-1/2})\), while recovery of the underlying
coefficient \(\boldsymbol{\beta}_0\) additionally reflects the finite-component
approximation error. For \(K=1\), both the oracle and estimated-weight
estimators admit projected asymptotic normality, with the estimated-weight
asymptotic variance capturing the covariance between the oracle and
weight-estimation contributions from the shared source sample alongside a
separate target-sample contribution. Both simulation studies—including
nonlinear response settings with a quadratic term—and real-data spectroscopic
applications demonstrate that IWPLS consistently improves target-domain
prediction over unweighted PLS and outperforms Ridge-IWLS under common
estimated weights.

Several directions merit further investigation. While
Supplementary Section~I.2 verifies the empirical weight-error
condition for correctly specified fixed-basis uLSIF under its stated
assumptions, extending the consistency and distributional theory to fitted PCA
representations and data-dependent tuning would bridge our theoretical
guarantees with the implemented PCA-uLSIF procedure. A second important
direction is to establish theoretical guarantees for IWCV-based component
selection and characterize the resulting estimation--approximation trade-off.

\bibliographystyle{plainnat}
\bibliography{references}

\end{document}